\documentclass[structabstract]{aa}  
\usepackage{natbib}
\bibpunct{(}{)}{;}{a}{}{,} 
\usepackage{graphicx}

\usepackage{txfonts}
\begin{document}

\title{GOODS-{\it Herschel}: the impact of galaxy-galaxy interactions 
  on the far-infrared properties of galaxies
}
\author{H. S. Hwang\inst{1}
\and D. Elbaz\inst{1}
\and M.~Dickinson\inst{2}
\and V. Charmandaris\inst{3,4}
\and E.~Daddi\inst{1}
\and D. Le Borgne\inst{6}
\and V. Buat\inst{5}
\and G. E. Magdis\inst{1,7}
\and B.~Altieri\inst{8}
\and H.~Aussel\inst{1}
\and D.~Coia\inst{8}
\and H.~Dannerbauer\inst{1}
\and K.~Dasyra\inst{1}
\and J.~Kartaltepe\inst{2}
\and R.~Leiton\inst{1}
\and B.~Magnelli\inst{9}
\and P.~Popesso\inst{9}
\and I.~Valtchanov\inst{8}
}
\institute{Laboratoire AIM-Paris-Saclay, CEA/DSM/Irfu - CNRS - Universit\'e Paris Diderot, CE-Saclay, F-91191 Gif-sur-Yvette, France \\
\email{hoseong.hwang@cea.fr}
\and National Optical Astronomy Observatory, 950 North Cherry Avenue, Tucson, AZ 85719, USA
\and Department of Physics and Institute of Theoretical \& Computational Physics, University of Crete, GR-71003 Heraklion, Greece
\and IESL/Foundation for Research and Technology - Hellas,  GR-71110, Heraklion, Greece and Chercheur Associ\'e, Observatoire de Paris, F-75014, Paris, France
\and Laboratoire d'Astrophysique de Marseille, OAMP, Universit´e Aixmarseille, CNRS, 38 rue Fr´ed´eric Joliot-Curie, 13388 Marseille cedex 13, France
\and Institut d'Astrophysique de Paris, UMR 7095, CNRS, UPMC Univ. Paris 06, 98bis boulevard Arago, F-75014 Paris, France
\and Department of Physics, University of Oxford, Keble Road, Oxford OX1 3RH, UK
\and Herschel Science Centre, European Space Astronomy Centre, Villanueva de la Ca\~nada, 28691 Madrid, Spain
\and Max-Planck-Institut f\"ur Extraterrestrische Physik (MPE), Postfach 1312, 85741, Garching, Germany
}

\date{Received September 15, 1996; accepted March 16, 1997}

   
\abstract
{}
{We study the impact of galaxy-galaxy interactions
  on the far-infrared properties of galaxies and its evolution at $0<z<1.2$.
}
{Using the high-$z$ galaxies in the fields of 
  Great Observatories Origins Deep Survey (GOODS) observed by Herschel Space Observatory
  in the framework of the GOODS-{\it Herschel} key program
  and the local IRAS or AKARI-selected galaxies 
  in the field of Sloan Digital Sky Survey Data Release 7,
  we investigate the dependence of galaxy properties
  on the morphology of and the distance to the nearest neighbor galaxy.
}
{We find that the star formation rates (SFRs) and 
  the specific SFRs (SSFRs) of galaxies, on average, 
  depend on the morphology of and the distance to the nearest neighbor galaxy
  in this redshift range. 
When a late-type galaxy has a close neighbor galaxy,
  the SFR and the SSFR increase as it approaches a late-type neighbor, 
  which is supported by Kolmogorov-Smirnov (K-S) and 
  Monte Carlo (MC) tests 
  with a significance level of $>$99\%.
However, the SFR and the SSFR
  decrease or do not change much as it approaches an early-type neighbor.
The bifurcations of SFRs and SSFRs depending on the neighbor's morphology
  seem to occur at $R_n\approx 0.5 r_{\rm vir,nei}$ 
  (virial radius of the neighbor),
  which is supported by K-S and MC tests with a significance level of $>$98\%. 
For all redshift bins, 
  the SSFRs of late-type galaxies interacting with late-type neighbors
  are increased by factors of about 1.8$\pm$0.7 and 4.0$\pm$1.2 compared to 
  those of non-interacting galaxies 
  when the pair separation is smaller than 
  $0.5 r_{\rm vir,nei}$ and $0.1 r_{\rm vir,nei}$, respectively.
The dust temperature
  of both local and high-$z$ late-type galaxies strongly interacting with late-type neighbors
  (i.e. $R_{\rm n} \leq0.1 r_{\rm vir,nei}$)
  appears to be higher than that of non-interacting galaxies
  with a significance level of $96-99$\%.
However,   
  the dust temperature
  of local late-type galaxies strongly interacting with early-type neighbors
  seems to be lower than or similar to that of non-interacting galaxies.
}
{Our results suggest that galaxy-galaxy interactions and mergers
  have been strongly affecting the SFR and the dust properties of star-forming galaxies
  over at least 8 billion years.}


\authorrunning{H. S. Hwang et al.}
\maketitle
%


\section{Introduction}

In the hierarchical picture of galaxy formation,
  galaxies are formed and grow through continuous
  interactions and mergers with other galaxies.
These galaxy-galaxy interactions and mergers
  are expected to strongly affect galaxy properties
  such as morphology, luminosity, 
  structure parameters, star formation rate (SFR),
  or dust properties 
  (see \citealt{str06} for a review).

Since the merger between spiral galaxies is known to form elliptical galaxies 
  as first suggested by \citet{too77},
  there is growing evidence for
  a change in galaxy morphology (e.g, \citealt{park08})
  and galaxy structure with merger 
  (e.g., \citealt{nik04, pat05, her05, pc09}).
For example, \citet{park08} showed that galaxy morphology and luminosity
  strongly depend on the distance to and 
  the morphology of the nearest neighbor galaxy.
When a galaxy is located within the virial radius of its nearest neighbor,
  its morphology tends to be the same as that of the neighbor.
This indicates an important role of hydrodynamical interactions with
  neighbors within the virial radius.
This morphological conformity was also found
  between host and their satellites galaxies \citep{ann08,wang10}, and
  between galaxies in galaxy clusters \citep{ph09}. 

For the star formation activity (SFA), 
  \citet{lt78} first noted that morphologically normal and peculiar galaxies
  show very different optical color distributions,
  which suggests that the SFA of peculiar galaxies is enhanced 
  compared to those of normal galaxies.
Many studies have extensively investigated this enhancement
  of SFA in paired galaxies
  (e.g., \citealt{con82, keel85, ken87, bar00, lam03, 
  nik04, alo04, 
  deprop05, gel06, woods07,
  li08, ell08, kna09, perez09, pc09, hwa10lirg, xu10, darg10, pat11}).
In summary, the SFA of galaxies seems to strongly depend on the morphology
  and the mutual separation of galaxies in pairs.
For example,
  the SFR of a galaxy is increased when the nearest neighbor has a late morphological type, 
  but is decreased or remains the same when the neighbor is an early-type galaxy 
  \citep{pc09,hwa10lirg,xu10}.
There are two characteristic pair-separation scales where 
  the SFA abruptly changes: 
  the virial radius ($r_{\rm vir,nei}$,
   to be defined in \S\ref{environ}) of the nearest neighbor galaxy 
  where the effects of galaxy interaction emerge
  (roughly a few hundred kpc for bright galaxies) and
  $\sim$0.05 $r_{\rm vir,nei}$ where the galaxies in pairs start to merge \citep{pc09}.
However, some other studies found no significant difference
  in galaxy properties between isolated and paired galaxies 
  (e.g., \citealt{ye95,pat97,ber03}).
For high-$z$ galaxies,
  the evolution of the fraction of interacting galaxies 
  has been studied extensively and is found to increase with redshift
  (e.g., \citealt{zk89, kar07, bri10}),
  but there are few studies focusing on the enhancement of SFA 
  in paired galaxies (e.g., \citealt{lin07,der09,woods10,wong11}).
 
For dust-enshrouded systems such as luminous infrared galaxies (LIRGs),
  one might have expected that
  not only the SFR but also
  the dust properties such as dust temperature or dust mass
  change as a function of pair separation.
Interestingly,
  \citet{tel88} found the highest dust color temperature 
  (i.e., the flux density ratio between {\it IRAS} 60 and 100 $\mu$m)
  for galaxies in pairs with the smallest separation.
On the other hand,
  \citet{lutz98} and \citet{rig99} did not find any correlation between
  the line-to-continuum ratio for the 7.7 $\mu$m PAH emission feature 
  (i.e. SF vs. AGN activity)
  and the nuclear separation of the interacting components of 
  ultraluminous infrared galaxies (ULIRGs). 
\citet{klaas01} also searched for the differences 
  in the mid-infrared (MIR), far-infrared (FIR) and submm 
  luminosity ratios of interacting and non-interacting ULIRGs,
  but did not find any noticeable difference depending on the merging stage.
However, \citet{xil04} found evidence for an increase
  in the 100- to 850-$\mu$m flux density ratio 
  (a proxy for the mass fraction of the warm and cold dust) 
  with the merging sequence.
They argue that the contradictory results
  may be caused by the different sample selection.
\citet{klaas01} did not order 
  their interacting ULIRG systems on the basis of the pair separation.
Their samples span a small range in bolometric luminosity,
  therefore, they might have missed some interacting systems 
  in the very early stage of merging.
It is also noted that \citet{smi07} found no significant change 
  in {\it Spitzer} MIR colors
  with pair separation in their sample of 
  tidally distorted premerger interacting galaxy pairs.
This may be because their sample was selected to be in the early stage of interactions.
  
Thanks to the advent of the {\it Herschel Space Observatory}  \citep{pil10} 
  with its very wide wavelength coverage ($70-500\mu$m),
  we are now able to have both the `Wien' and `Rayleigh-Jeans' sides of 
  the FIR spectral energy distributions (SEDs) of high-$z$ galaxies,
  and to have robust SFR and dust temperature measurements for them
  (e.g., \citealt{elb10,elb11,hwa10tdust,gmag10bumpy,gmag10lbg}).
In this paper, 
  we investigate the impact of galaxy-galaxy interactions
  on galaxy properties focusing on FIR properties
  of high-$z$ galaxies at $0.4 \la z \la 1.2$
  by taking advantage of the wide wavelength coverage
  of the Photodetector Array Camera (PACS; \citealt{pog10}) and 
  Spectral and Photometric Imaging Receiver (SPIRE; \citealt{gri10}) 
  instruments onboard {\it Herschel}
  in the fields of Great Observatories Origins Deep Survey 
  (GOODS; \citealt{dic03,gia04}).
To compare these results with those of local galaxies,
  we also use a sample of galaxies at $z<0.1$ that were covered by
  {\it IRAS} and {\it AKARI} satellites \citep{mur07} in 
  the field of Sloan Digital Sky Survey (SDSS, \citealt{york00}). 
The {\it IRAS} and {\it AKARI} all-sky survey data 
  contain flux density measurements at 12$-$160 $\mu$m,
  which can probe both the `Wien' and `Rayleigh-Jeans' sides of 
  the peak of IR SEDs 
  of local galaxies in a way similar to that of {\it Herschel} 
  for high-$z$ galaxies. 
Section \ref{data} describes the data used in this study,
  and the dependence of galaxy properties 
  on the distance to and the morphology of the nearest neighbor galaxies 
  is given in \S \ref{results}.
Discussion and conclusions are given in 
  \S \ref{discuss} and \S \ref{con}, respectively.
Throughout, we adopt $h=0.7$ and a flat $\Lambda$CDM cosmology 
  with density parameters 
  $\Omega_{\Lambda,0}=0.73$ and $\Omega_{m,0}=0.27$.

\section{Data}\label{data}
\subsection{GOODS Sample}\label{gdata}

We used a spectroscopic sample of galaxies in GOODS,
  which is a deep multiwavelength survey covering two carefully selected regions
  including the Hubble Deep Field North (HDF-N) and 
  the Chandra Deep Field South (CDF-S).
Hereafter, the two GOODS fields centered on HDF-N and CDF-S are called
  GOODS-N and GOODS-S, respectively.
The combined area of the two fields is approximately 320 arcmin$^2$.

The GOODS fields were observed by {\it Herschel}
  in the GOODS-{\it Herschel} key program \citep{elb11}.
The full $10\arcmin\times16\arcmin$ GOODS-N field
  was observed with both PACS (100 and 160 $\mu$m) and 
  SPIRE (250, 350, and 500 $\mu$m).
A smaller region within the GOODS-S field 
   ($\approx 10\arcmin\times10\arcmin$) was observed with PACS only.
We supplemented these data with public SPIRE images 
  covering the full $10\arcmin\times16\arcmin$ GOODS-S field,
  which was originally taken in the HerMES key program (Oliver et al. in prep.).
Source extraction on these PACS and SPIRE images 
  was performed at the prior positions of {\it Spitzer} 24 $\mu$m-selected sources, 
  and details are described in \citet{elb11}.
Note that this extraction method with 24 $\mu$m-selected sources
  recovers more than 99\% of {\it Herschel} sources \citep{gmag11}.
We used flux densities in PACS bands down to 3$\sigma$ limits of 
  1.1 and 2.7 mJy (0.8 and 2.4 mJy) at 100 and 160 $\mu$m in GOODS-N
  (GOODS-S), respectively.  
SPIRE measurements are used down to 5$\sigma$ limits of
  5.7, 7.2, and 9.0 mJy at 250, 350 and 500 $\mu$m, respectively
  (see Table 1 in \citealt{elb11} for more details about the noise properties).
By combining {\it Herschel} data with the existing multi-wavelength data, 
  we made a band-merged catalog of GOODS galaxies
  using the photometric data
  at {\it HST} ACS $BViz$,
  CFHT/WIRCam $JK$ (North; \citealt{wang10wir}) and 
  VLT/ISAAC $JHK$ (South; \citealt{ret10}),
  {\it Spitzer} IRAC 3.6, 4.5, 5.8, 8 $\mu$m and MIPS 24 and 70 $\mu$m, 
  {\it Herschel} PACS 100 and 160 $\mu$m, and
  {\it Herschel} SPIRE 250, 350 and 500 $\mu$m.   

\begin{figure*}
\center
\includegraphics[scale=0.5]{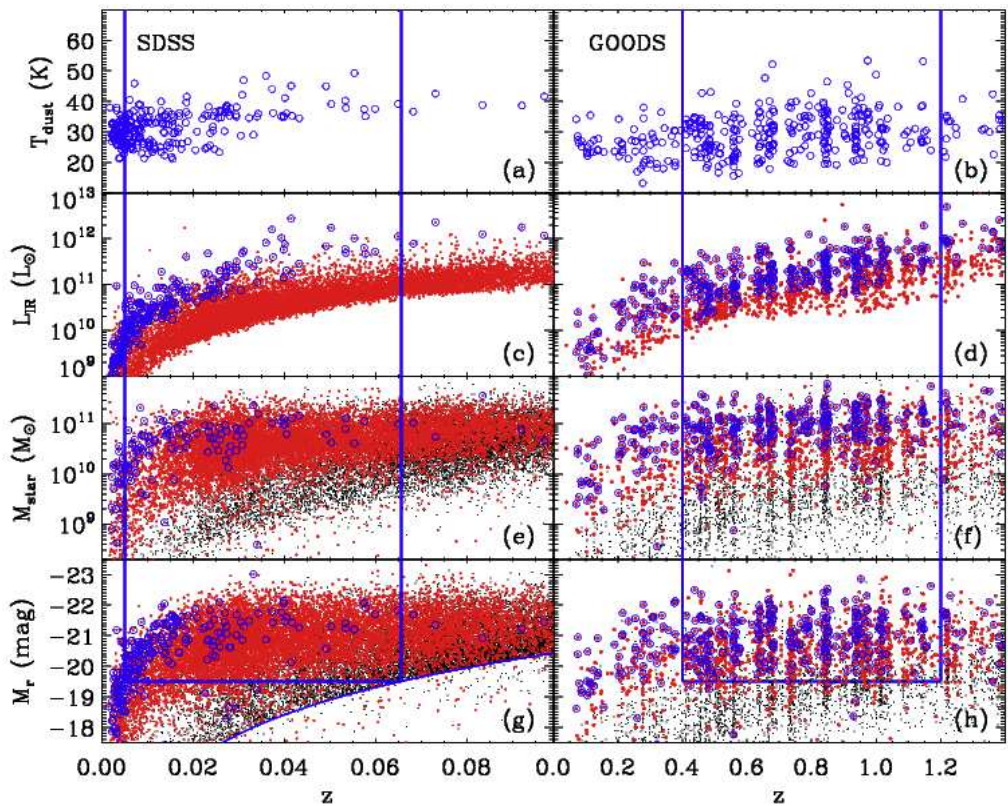}
\caption{(a-b) Dust temperature ($T_{\rm dust}$), 
  (c-d) IR luminosity, (e-f) stellar mass, 
  and (g-h) evolution-corrected, rest frame $r$-band absolute magnitude 
  vs. spectroscopic redshift
  for galaxies in ({\it left}) SDSS and ({\it right}) GOODS.
Red dots indicate FIR-selected galaxies.
Black dots denote
  galaxies without FIR detection in the spectroscopic sample of galaxies
  (only 3 $\%$ of SDSS galaxies in the total sample are shown).
Open blue circles denote galaxies with $T_{\rm dust}$ measurements.
Blue solid lines in (g-h) define the volume limited samples of 
  SDSS galaxies at $0.005\leq z\leq0.0656$
  and of GOODS galaxies at $0.4\leq z\leq 1.2$, respectively.
The bottom curve in (g) indicates the apparent magnitude limit ($m_r=17.77$)
  for the main galaxy sample in SDSS 
  using the mean $K$-correction relation given by Eq. (2) of \citet{choi07}.
 }\label{fig-vol}
\end{figure*}

Among the sources in the band-merged catalog, 
  we used 3630 and 3181 galaxies whose spectroscopic redshifts are reliable
  over the whole fields of
  GOODS-N (\citealt{coh00,cow04,wir04,red06,bar08,coo11}; Stern et al. in prep) and 
  GOODS-S \citep{szo04,lef04,mig05,van05,van06,van08,rav07,pop09,kurk09,bal10,sil10,xia11},
  respectively.
The rest frame $r$-band absolute magnitude $M_r$ of galaxies 
  was computed based on the ACS plus near-infrared (NIR) photometry
  with 
  $K$-corrections \citep{bla07kcorr}.
It was computed in fixed bandpasses, shifted to $z=0.1$
  to be compared with local SDSS galaxies.
The $1.1(z-0.1)$ term was added 
  to $M_r$ for the evolution correction \citep{wolf03}.
We calculated stellar masses using the photometric data
  upto IRAC 4.5 $\mu$m with a Salpeter IMF \citep{sal55}. 
We used the code Z-PEG
  with a galaxy evolution model P{\'E}GASE.2 
  (\citealt{fr99,lr02}; see \citealt{elb11} for details).

We adopted the galaxy morphology information [early types (E/S0) and late types (S/Irr)]
  from \citet{hp09} 
  that is based on the visual inspection of ACS $BViz$ images.
We performed additional visual classification 
  for the galaxies in the volume-limited sample of galaxies 
  shown in the right panels of Fig. \ref{fig-vol}
  that are not included in \citet{hp09}.
  
We computed the IR luminosity ($L_{\rm IR}$) for 903 and 828 galaxies
  with spectroscopic redshifts
  in GOODS-N and -S, respectively,
  detected in at least one out of the five PACS/SPIRE bands.
We fit the flux densities at $\lambda_{\rm rest}\geq30\mu$m
  by allowing the normalization of the SED templates of \citet[CE01]{ce01}
  to vary and choosing the one that minimizes the $\chi^2$ values.
When there are two or less data points to fit (i.e. $N\leq2$),
  we fit the flux densities without allowing the
  normalization of the templates (i.e. standard CE01 technique).
There are some cases with only one FIR band used for $L_{\rm IR}$ measurement,
  but the IR luminosities extrapolated from a single passband
  were found to agree very well with those measured with all FIR bands 
   with an average uncertainty of $\sim$30\% (see Fig. 23 of \citealt{elb11}).
Therefore, this does not introduce any bias in our results.
To determine the dust temperature,
  we fit the observational data with a modified black body (MBB) model by fixing
  the emissivity parameter to $\beta=1.5$.
We require at least one flux measurement
  at each `Wien' and `Rayleigh-Jeans' side of the FIR peak
  [i.e., at least two measurements in total; 
  see \citealt{hwa10tdust} for detailed selection criteria].
In the result, we have 284 and 104 galaxies with $T_{\rm dust}$ measurements
  in GOODS-N and -S, respectively.

\subsection{Sloan Digital Sky Survey Sample}\label{sdss}

For local galaxies, 
  we used a spectroscopic sample of galaxies in 
  the SDSS Data Release 7 \citep[SDSS DR7]{aba09} 
  complemented by a photometric sample of SDSS galaxies
  whose redshift information is not available in the SDSS database,
  but available in the literature \citep{hwa10lirg}.
In addition, we used the FIR data for these galaxies
  compiled in \citet{hwa10tdust}
  by cross-correlating 
  {\it IRAS} Faint Sources Catalog -- Version 2 (\citealt{mos92})
  and {\it AKARI}/Far-Infrared Surveyor (FIS; \citealt{kaw07})
  all-sky survey Bright Source Catalogue
  (BSC\footnote{http://www.ir.isas.jaxa.jp/AKARI/Observation/PSC/ 
  Public/RN/AKARI-FIS$\_$BSC$\_$V1$\_$RN.pdf}) ver. 1.0 
  with the SDSS samples.

The $r$-band absolute magnitude $M_r$ was also computed 
  in fixed bandpasses, shifted to $z=0.1$,
  using Galactic reddening correction \citep{sch98} and $K$-corrections \citep{bla07kcorr}.
The evolution correction given by \citet{teg04}, $E(z) = 1.6(z-0.1)$, 
  is also applied.
Note that the amount of evolution correction is different between GOODS and SDSS galaxies.
These values are taken from the redshift evolution of the characteristic luminosity 
  in the luminosity function of local and high-z galaxies separately,
  so we kept different values.
Change of these values does not affect our conclusions.

By adopting a method similar to the one applied to GOODS galaxies,
  we computed the IR luminosity of 14444 galaxies
 (among the total sample of 926,748 SDSS galaxies)
  whose {\it IRAS} 60 $\mu$m or {\it AKARI} 90 $\mu$m flux densities are 
  reliable\footnote{Flux quality flags 
  are either `high' or `moderate' for {\it IRAS} sources
  and `high' for {\it AKARI} sources.} 
  using the CE01 SED templates
  by allowing normalization of the templates.
We fit the flux densities without allowing the
  normalization of the templates 
  if there are two or less data points to fit.
To determine the dust temperatures,
  we again fit the observational data with a modified black body (MBB) model 
  by fixing the emissivity parameter to $\beta=1.5$ 
  only for 238 galaxies detected at {\it AKARI} 140 or 160 $\mu$m
  so that we can have flux density measurements longwards of the FIR peak
  as well as the one shortwards of the peak
  in a similar way to GOODS-{\it Herschel} galaxies
  (see \citealt{hwa10tdust} for detailed selection criteria).

The stellar mass estimates were obtained from 
  MPA/JHU DR7 value-added galaxy catalog\footnote{http://www.mpa-garching.mpg.de/SDSS/DR7/}
  (VAGC), which are based on the fit of SDSS five-band photometry 
   \citep{kau03,gal05}.
We convert these estimates based on Kroupa IMF \citep{kro01}
  to a Salpeter IMF by dividing them by a factor of 0.7 \citep{elb07}.

We adopted the galaxy morphology information from
  the Korea Institute for Advanced Study (KIAS) DR7 VAGC\footnote{http://astro.kias.re.kr/vagc/dr7/} 
 \citep{pc05,choi10}.
We performed additional visual classification 
  for the galaxies in the SDSS database 
  that are not included in KIAS DR7 VAGC.

\subsection{Comparison of GOODS and SDSS galaxies}\label{tdust}
 
In the right panels of Fig. \ref{fig-vol}, 
  we plot several physical parameters of the GOODS galaxies
  as a function of redshift, 
  and define a volume limited sample to be analyzed
  ($-19.5\geq M_r$ with $0.4\leq z\leq1.2$).
Similarly, we plot the SDSS galaxies in the left panels of Fig. \ref{fig-vol},
  and define a volume limited sample of galaxies with 
  $-19.5\geq M_r$ and $0.005\leq z\leq0.0656$.
The comoving volume for this SDSS sample is $\sim2.0\times10^7$ Mpc$^{3}$,
  which is much larger than that for
  the GOODS galaxies ($\sim$2.4$\times10^5$ Mpc$^{3}$
  for GOODS-N at $0.4\leq z\leq 1.2$).
Note that the FIR detection limits for local SDSS and high-$z$ GOODS galaxies 
  are not the same even if we fix the mass and luminosity of galaxies in \S\ref{results}.
However, since we are interested in the difference of SFA
  depending on the morphology of and the distance to the nearest neighbor galaxy
  in a given redshift range,
  the different FIR detection limits between local and high-$z$ galaxies
  do not affect our conclusions.

\begin{figure}
\center
\includegraphics[scale=0.7]{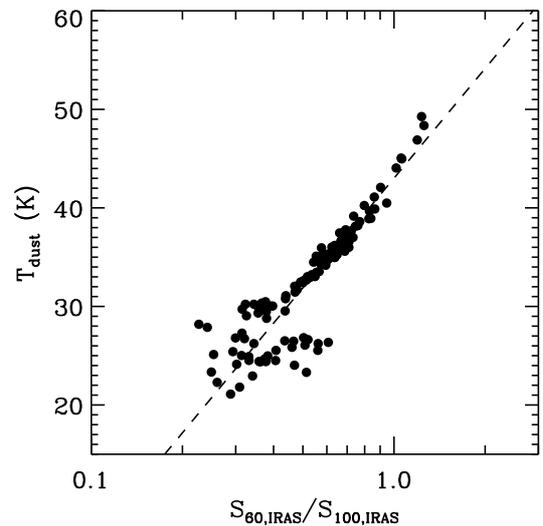}
\caption{Correlation between
  dust temperature ($T_{\rm dust}$)
  and flux density ratios  ($S_{\rm 60,IRAS}/S_{\rm 100,IRAS}$) for local SDSS galaxies.
  The dashed line represents the best-fit with an ordinary least-squares bisector method.
}\label{fig-fluxtdust}
\end{figure}

In Fig. \ref{fig-fluxtdust},
  we plot the dust temperature versus FIR flux density ratio ($S_{\rm 60,IRAS}/S_{\rm 100,IRAS}$)
  for SDSS galaxies,
  which shows a good correlation between the two quantities (see also \citealt{cha07}).
From the fit with an ordinary least-squares bisector method \citep{iso90},
  we derive a transformation relation between the two quantities,
     
\begin{equation}
T_{\rm dust}~({\rm K})=(43.0\pm 0.3) + (37.0\pm 1.5){\rm log}(S_{\rm 60,IRAS}/S_{\rm 100,IRAS}).
\label{eq-tdust}
\end{equation}

Since the number of SDSS galaxies with $S_{\rm 60,IRAS}/S_{\rm 100,IRAS}$
  is larger than that with $T_{\rm dust}$ measurements,
  we transform the FIR flux density ratio into the dust temperature using this equation
  to increase the statistics (see Fig. \ref{fig-tdust}).
   

\subsection {Nearest neighbor galaxy}\label{environ}

To investigate the effects of interactions with the nearest neighbor galaxy,
  we first identified the nearest neighbor of a target galaxy
  that is the closest to the target galaxy on the projected sky
  and that satisfies the conditions of magnitude and relative velocity.
We searched for the nearest neighbor galaxy among galaxies
  that have magnitudes brighter than $M_r=M_{r,\rm target}+0.5$ and 
       have relative velocities less than 
       $\Delta \upsilon=|\upsilon_{\rm neighbors}- \upsilon_{\rm target}|=660$ km s$^{-1}$
       for early-type target galaxies and less than $\Delta \upsilon=440$ km s$^{-1}$ 
       for late-type target galaxies.
Because the redshift uncertainties are larger for the GOODS galaxies, 
  we use velocity difference limits that are 10\% larger than 
  those that we have allowed for SDSS galaxies 
  in previous studies
  ($\Delta \upsilon=600$ and 400 km s$^{-1}$
   for early- and late-type target SDSS galaxies, 
   respectively, as seen in Fig. 1 of \citealt{park08}).  
The use of different values for the relative velocity condition depending on
  galaxy morphology is supported by
  different velocity distributions of neighboring galaxies
  depending on galaxy morphology,
  as shown in Fig. 1 of \citet{park08} and Fig. 2 of \citet{pc09} for SDSS galaxies
  and Fig. 3 of \citet{hp09} for GOODS galaxies.
Since we used volume-limited samples of galaxies with $M_r\leq-19.5$,
  we restrict our analysis to target galaxies 
  brighter than $M_{r,\rm target}=-20$ so that their neighbors 
  ($M_{r,\rm nei}\leq M_{r,\rm target}+0.5$)
  are searched within the volume-limited samples.

The virial radius of a galaxy within which 
  the mean mass density is 200 times the critical density 
  of the universe ($\rho_c$), is calculated by
\begin{equation}
r_{\rm vir}=(3 \gamma L/4\pi)^{1/3} (200\rho_c)^{-1/3},
\label{eq-vir}
\end{equation}
where $L$ is the galaxy luminosity, and $\gamma$ the mass-to-light ratio.
Here, the mass associated with a galaxy plus dark halo
  system is assumed to be proportional to the $r$-band luminosity of the galaxy.
We assume that the mass-to-light ratio of early-type galaxies
  is on average twice as large as that of late-type galaxies
  at the same absolute magnitude $M_r$,
  which means $\gamma$(early)$=2\gamma$(late) 
 [see \S 2.5 of \citet{pc09} and \S 2.4 of \citet{park08} for SDSS galaxies,
  and \S 2.3 of \citet{hp09} for GOODS galaxies].
The critical density of the universe ${\rho}_c$ 
  is a function of redshift $z$
  [i.e. ${\rho}_c=3H^2(z)/(8\pi G)$] and 
 $\Omega_m(z)= \rho_b(z)/\rho_c (z)=\overline{\rho}(1+z)^3/\rho_c(z)$,
 where $\rho_b$ and $\overline{\rho}$ are the mean matter densities 
 in proper and comoving spaces, 
 respectively\footnote{We correct a typo for $\Omega_m(z)$ in \citet{hp09}, 
 but all the related values in that paper are correct.}.
The Hubble parameter at $z$ is
  $H^2(z)=H^2_0 [\Omega_{m,0}(1+z)^3 +\Omega_{k,0}(1+z)^2+\Omega_{\Lambda,0}]$,
  where $\Omega_{m,0}$, $\Omega_{k,0}$, and $\Omega_{\Lambda,0}$ 
  are the dimensionless density parameters at the present epoch \citep{pee93}.
Then, the virial radius of a galaxy at redshift $z$ in proper space 
  can be rewritten by
\begin{equation}
r_{\rm vir} (z) = [3 \gamma L \Omega_{m,0} / (800\pi \overline{\rho}) / \{ \Omega_{m,0}(1+z)^3 + \Omega_{k,0}(1+z)^2 + \Omega_{\Lambda,0} \} ]^{1/3}.
\label{eq-vir2}
\end{equation}

The value of mean density of the universe for local SDSS galaxies, 
  $\bar\rho=(0.0223\pm0.0005)(\gamma L)_{-20} (h^{-1}{\rm Mpc})^{-3}$,
  was adopted where $(\gamma L)_{-20}$ is the mass of a late-type galaxy 
  with $M_r=-20$ \citep{park08}.

For high-$z$ GOODS galaxies,
  we computed the mean mass density $\overline{\rho}$ using
  the galaxies at $z=0.4-1.2$ with
  various absolute magnitude limits varying from $M_r= -16$ to $-20$.
We found that the mean mass density
  appears to converge when the magnitude cut is fainter than $M_r=-17.5$,
  which means that the contribution of faint galaxies
  is not significant because of their small masses.
In this calculation,
  each galaxy is weighted by the inverse of completeness
  according to its apparent magnitude and color (see Fig. 1 of \citealt{hp09}).
We obtain $\overline{\rho} =$ 0.017 and 0.013 $ (\gamma L)_{-20}$ (Mpc$^{-3}$)
  for GOODS-N and -S, respectively,
  where $(\gamma L)_{-20}$ is the mass of a late-type galaxy with $M_r=-20$.
According to our formula the virial radii of galaxies with
  $M_r=-20$ and $-21$ are 300  and 400 $h^{-1}$ kpc for early types,
  and 240 and 320 $h^{-1}$ kpc for late types, respectively.  

The spectroscopic completeness can affect the identification of 
  the genuine nearest neighbor, and then the nearest neighbor 
  can be seriously misidentified if the completeness is very low. 
Our previous Monte Carlo experiment shows that the fraction of 
  the misidentified nearest neighbor reaches about 50\% 
  when the sample completeness is 50\% \citep{hp09}. 
Therefore, it is necessary to have survey data with high completeness 
  in order not to miss the genuine nearest neighbor. 
Up to now, GOODS has the highest spectroscopic completeness 
  ($71-86\%$ for GOODS galaxies at $m_z<23$)
  among several large, deep-field surveys with {\it HST} images
  to our knowledge.
Therefore, it is the best survey data for our analysis,
  but it should be noted that our results 
  could be weakened by this incompleteness. 
The completeness depends on the apparent magnitude and color 
  (see Fig. 1 of \citealt{hp09}) and also depends on the distance between galaxies 
  due to the difficulty in observing galaxies close to each other 
  using multiobject spectrograph (MOS). 
We checked the completeness as a function of the projected distance to the target galaxy, 
  and found that it does not change with the projected distance. 
It might be because we combined spectroscopic data from numerous references, 
  therefore, the difficulty in observing galaxies 
  with small separation using MOS is significantly reduced.
Similarly, the redshift information of some SDSS galaxies missed by the SDSS database
  was complemented by the data in the literature \citep{hwa10lirg},
  so there is also no bias for local galaxies.

Note that the nominal FWHMs of the point spread function (PSF)
  are 37\arcsec (65 $\mu$m), 39\arcsec (90 $\mu$m), 
     58\arcsec (140 $\mu$m), and 61\arcsec (160 $\mu$m)
  for {\it AKARI} bands \citep{kaw07},
  and 1.44\arcmin (60 $\mu$m) and 2.94\arcmin (100 $\mu$m) 
  for {\it IRAS} bands \citep{san03}.
The corresponding angular size of one viral radius
  for typical late- and early-type galaxies with $M_r=-21$ and $z=0.035$,
  is $11.3\arcmin$ and $14.3\arcmin$, respectively.
Therefore, though the galaxies in pairs are selected from
  galaxy catalogs in optical bands with high spatial resolutions,
  the measured SFRs of galaxies in close pairs cannot be clearly 
  assigned to the individual galaxies 
  due to the poor spatial resolution of FIR data.
For example, 
  local galaxies with $M_r=-21$ and $z=0.035$
  having neighbors at $\lesssim0.06r_{\rm vir}$ ($\lesssim0.12r_{\rm vir}$)
  are not resolved with {\it AKARI} 90 $\mu$m ({\it IRAS} 60 $\mu$m),
  because the pair separation is smaller than the FWHM at each band.
Thus the measured SFRs for these galaxies can indeed indicate SFRs of 
  the whole interacting systems.
However, our results on the increased SFRs (or SSFRs) due to galaxy-galaxy interactions
  to be seen in \S\ref{results},
  are not strongly affected by this effect
  because the increased SFRs (or SSFRs) 
  are found to be much larger than a factor of two
  that could be simply due to this blending problem of two galaxies   
  in one FIR beam.

For high-$z$ galaxies, the PSF FWHMs 
  are 6.0$\arcsec$ ({\it Spitzer} 24 $\mu$m), 
  6.7$\arcsec$ ({\it Herschel} 100 $\mu$m), 
  11.0$\arcsec$ (160 $\mu$m), 
  18.1$\arcsec$ (250 $\mu$m), 
  24.9$\arcsec$ (350 $\mu$m), and
  36.6$\arcsec$ (500 $\mu$m).
The angular size of one virial radius
  for typical late- and early-type galaxies with $M_r=-21$
  would be, respectively, 70.8$\arcsec$ and 88.6$\arcsec$ at $z=0.6$, and
  59.1$\arcsec$ and 73.5$\arcsec$ at $z=1.0$.
Similar to the case of local galaxies,
  high-$z$ galaxies having neighbors at $\lesssim0.07r_{\rm vir}$ ($\lesssim0.09r_{\rm vir}$)
  are not resolved with {\it Spitzer} 24 $\mu$m
  (source extraction on {\it Herschel} images was performed 
  at the prior positions of {\it Spitzer} 24 $\mu$m-selected sources; 
  \citealt{elb11}).
However, this has no effect on our conclusions.

\begin{figure*}
\center
\includegraphics[scale=0.65]{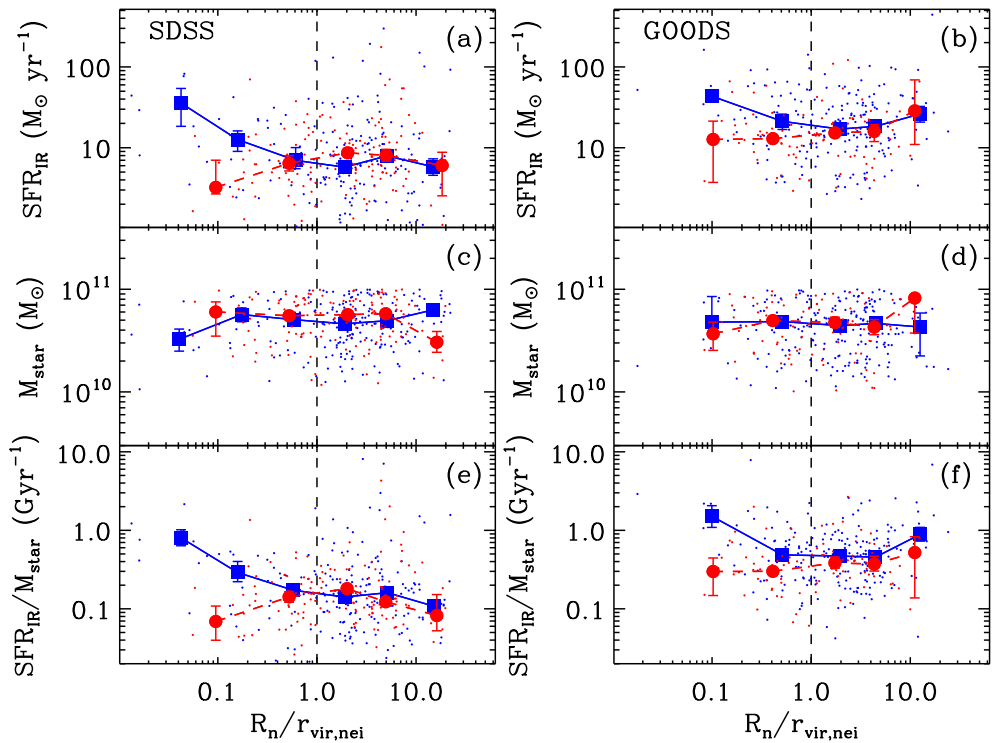}
\caption{
(a-b) SFR, (c-d) M$_{\rm star}$, (e-f) specific SFR 
  for late-type galaxies with $-20\geq M_r$ and 
  $10^{10}\leq$ M$_{\rm star}$ (M$_{\odot}$) $\leq 10^{11}$
  as a function of the projected distance 
  to the nearest neighbor galaxy ($R_{\rm n}/r_{\rm vir,nei}$)
  in ({\it left}) SDSS ($0.005\leq z\leq0.0656$) and
  ({\it right}) GOODS ($0.4\leq z_{\rm spec}\leq 1.2$).
Red and blue dots are late-type galaxies having early- and late-type 
  nearest neighbor galaxies, respectively. 
Large circle (early-type neighbor case) and square (late-type neighbor case)
  are median values of each physical parameter
  and of $R_{\rm n}/r_{\rm vir,nei}$ at each distance bin.
The errorbars represent $68\%$ $(1\sigma)$ confidence intervals 
  that are determined by the bootstrap resampling method.
}\label{fig-neibl}
\end{figure*}

\section{Results}\label{results}
\subsection{Change in SSFRs as a function of pair separation}

In Fig. \ref{fig-neibl},
  we plot several FIR properties of local (left) and high-$z$ (right) 
  late-type galaxies
  as a function of the projected distance 
  to the nearest neighbor galaxy
  normalized by the virial radius
  of the neighbor ($R_{\rm n}/r_{\rm vir,nei}$). 
Note that we only plot {\it IRAS} or {\it AKARI}-selected (left)
  and {\it Herschel}-selected (right) late-type galaxies,
  but their nearest neighbors are selected 
  among the spectroscopic samples of SDSS and GOODS galaxies regardless of 
  FIR detections.
To remain complete in terms of neighbor galaxies
  and remove the effect of mass on the SFA, 
  we restrict our analysis
  to the late-type galaxies with $-20\geq M_r$ and 
  $10^{10}\leq$ M$_{\rm star}$ (M$_{\odot}$) $\leq 10^{11}$.
Therefore, even if we use the physical distance
  for the pair separation without using a normalization by the virial radius,
  the observed trends in this figure do not change.

For high-$z$ samples,
  we only plot GOODS galaxies that
  have at most one neighbor within $6\arcsec$
  (the full width half maximum, FWHM, of the {\it Spitzer} beam at 24 $\mu$m)
  with $S_{24}>50\%$ of the central 24 $\mu$m source
  (i.e. $N_{\rm nei, 24\mu m}\leq1$).
This criterion is introduced 
  to reduce the contamination of neighboring sources,
  but to allow for the possibility of having one close neighbor 
  in order to study the effects of galaxy proximity. 
In practice, this criterion results in the removal of only 4 galaxies among 330 GOODS galaxies 
  in Fig. \ref{fig-neibl}, which has no effect on our conclusions.

Assuming a Salpeter IMF, 
  we converted the IR luminosity into SFR$_{\rm IR}$ 
  using the relation in \citet{ken98}: 
  SFR$_{\rm IR}$ ($M_\odot$ yr$^{-1}$) $= 1.72\times10^{-10}L_{\rm IR} (L_\odot)$.
Panels (a-b) show that SFR$_{\rm IR}$ depends on 
  the projected distance to the 
  nearest neighbor as well as on the neighbor's morphology.
Large circle (early-type neighbor case) and square (late-type neighbor case)
  are median values of each physical parameter
  and of $R_{\rm n}/r_{\rm vir,nei}$ at each distance bin.
It is seen that SFR$_{\rm IR}$ changes with neither the distance to the neighbor
  nor the neighbor's morphology
  when a galaxy is located farther than 0.5 $r_{\rm vir,nei}$ 
  (virial radius of neighbor galaxy).
On the other hand, when a galaxy is located 
  at $R_{\rm n} \lesssim 0.5 r_{\rm vir,nei}$,
  the SFR$_{\rm IR}$ increases 
  as the target, late-type galaxy approaches a late-type neighbor, but
  decreases or remains constant as it approaches an early-type neighbor.
It is important to note that the bifurcation of SFR$_{\rm IR}$
  depending on the neighbor's morphology
  occurs at $R_{\rm n}\approx 0.5 r_{\rm vir,nei}$.

\begin{figure}
\center
\includegraphics[scale=0.75]{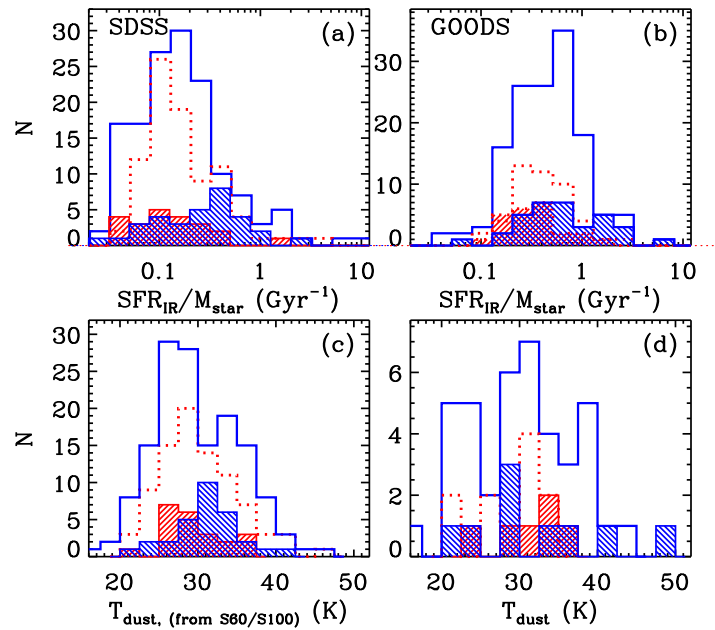}
\caption{Distribution of SSFRs ({\it top}) and $T_{\rm dust}$ ({\it bottom})
  for SDSS ({\it left}) and GOODS ({\it right}) galaxies.
Non-interacting galaxies 
  (i.e. $R_{\rm n} > r_{\rm vir,nei}$ in Fig. \ref{fig-neibl})
  with early- and late-type neighbors
  are shown by solid and dotted histograms, respectively.
Interacting galaxies (i.e. $R_{\rm n} \leq0.5 r_{\rm vir,nei}$)
  with early- and late-type neighbors
  are denoted by hatched histograms with
  orientation of 45$^\circ$ ($//$ with red color) and 
  of 315$^\circ$ ($\setminus\setminus$ with blue color) relative to horizontal, respectively.
}\label{fig-hist}
\end{figure}

\begin{figure*}
\center
\includegraphics[scale=0.65]{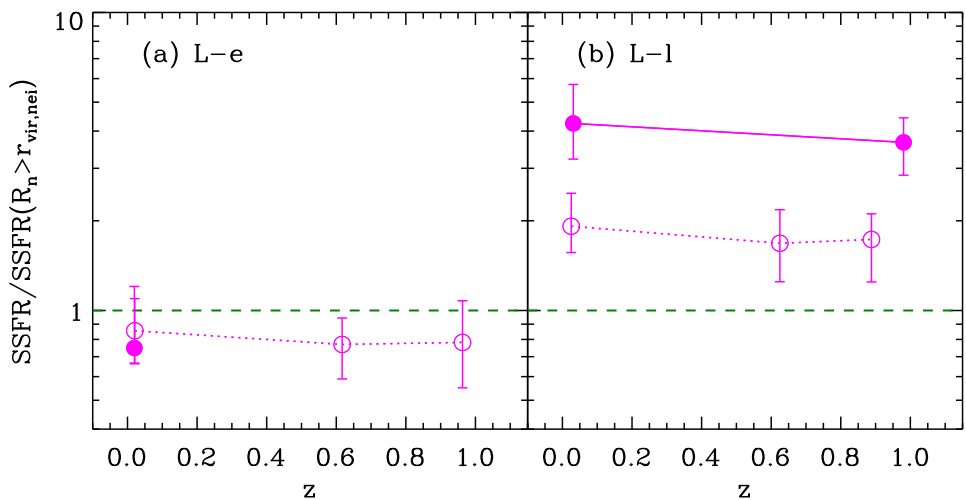}
\caption{SSFRs of interacting galaxies normalized by SSFRs
  of non-interacting galaxies 
  (i.e. $R_{\rm n} > r_{\rm vir,nei}$ in Fig. \ref{fig-neibl})
  as a function of redshift for a) early-type and b) late-type neighbor case.
Filled and open circles are median values for 
  strongly- (i.e. $R_{\rm n} \leq0.1 r_{\rm vir,nei}$),
  and weakly-interacting 
  (i.e. $R_{\rm n} \leq0.5 r_{\rm vir,nei}$) galaxies, respectively,
  at each redshift bin 
  ($0.005\leq z\leq0.0656$, $0.4\leq z\leq 0.8$, and $0.8\leq z\leq 1.2$).
}\label{fig-zevol}
\end{figure*}

Since the SFR is strongly correlated with the mass 
  \citep{bri04,elb07,dad07,gmag10sfrm},
  it is important to check the effects of galaxy mass on the change of SFR.
Thus we plot the distributions of stellar masses and
  specific SFRs (SFR$_{\rm IR}$/M$_{\rm star}$, SSFRs) in panels (c-f).
Panels (c-d) show no significant change of masses as a function of $R_{\rm n}$,
  but panels (e-f) suggests that a difference in SSFRs between early- and late-type neighbor cases 
  is still prominent at $R_{\rm n} \lesssim 0.5 r_{\rm vir,nei}$.
We also checked the redshift distributions 
  as a function of $R_{\rm n}$ (not shown), which 
  shows again no significant dependence on $R_{\rm n}$ and the morphology of neighbor.
We used a Kolmogorov-Smirnov (K-S) test to determine whether 
  the SSFR distributions of
  interacting and non-interacting galaxies 
  with late-type neighbors
  are drawn from the same distribution.
We tested two cases of interacting galaxies with late-type neighbors :
  1) strongly-interacting ($R_{\rm n} \leq0.1 r_{\rm vir,nei}$) galaxies versus
  non-interacting galaxies ($R_{\rm n} > r_{\rm vir,nei}$),
  and 2) relatively weakly-interacting ($R_{\rm n} \leq0.5 r_{\rm vir,nei}$) galaxies versus
  non-interacting galaxies.
The hypothesis that the two distributions are extracted
  from the same parent population can be rejected at the confidence level of 
  $>$99\% for both cases in the local universe.
In the high-$z$ universe,
  the hypothesis can be rejected at the confidence level of 
  $>$99 and 90\% for 
  strongly- and weakly-interacting galaxies, respectively.
To help the understanding of the difference in SSFRs for several subsamples,
  we plot, in Fig. \ref{fig-hist}, the distribution of SSFRs
  for interacting and non-interacting  galaxies in SDSS and GOODS.

The statistical significance of the increased SSFRs of 
  interacting galaxies with late-type neighbors
  compared to non-interacting galaxies,
  is also tested by Monte Carlo (MC) test.
We constructed two subsamples 
  (with the same number of galaxies as in the actual samples)
  by randomly drawing the SSFRs from the whole galaxy sample
  and computed the median of each subsample.
The resulting two subsamples will have the same medians on average.
These random subsamples will tell us about whether or not
the difference in SSFRs between
  interacting and non-interacting galaxies with late-type neighbors
  is statistically significant. 
We constructed 1000 trial data sets and 
  computed the fraction of simulated data sets 
  in which the difference of the SSFRs is larger than 
  or equal to that based on the real data (f$_{{\rm sim}\geq {\rm obs}}$).
The significance levels of the difference defined by 100(1$-$f$_{{\rm sim}\geq {\rm obs}}$) (\%).
This test also confirms the increased SSFRs of 
  interacting galaxies with late-type neighbors
  compared to non-interacting galaxies
  with a significance level of $>$99\%
  for both strongly- and weakly-interacting galaxies 
  in local and high-$z$ universe.

\begin{figure*}
\center
\includegraphics[scale=0.7]{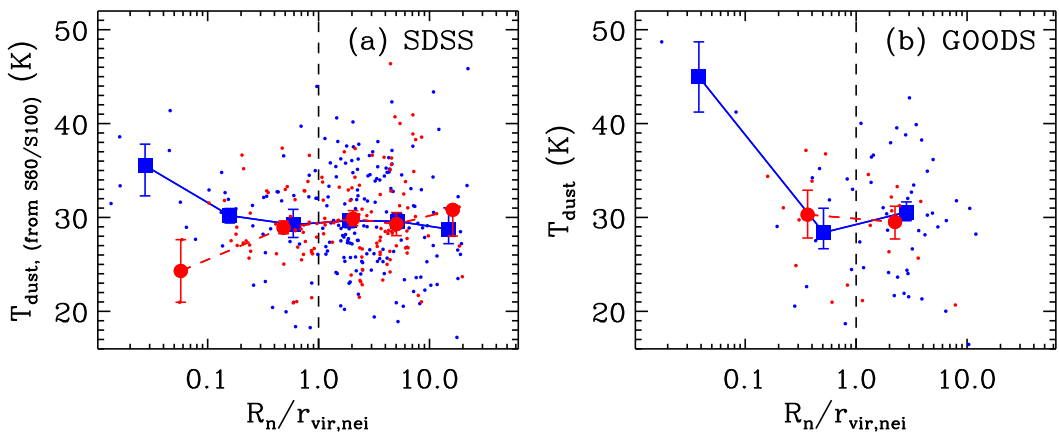}
\caption{Same as Fig. \ref{fig-neibl}, but for 
  the dust temperature of (a) SDSS and (b) GOODS galaxies.
}\label{fig-tdust}
\end{figure*}

Similarly, the statistical significance of different SSFR distributions of interacting galaxies
  with early- and late-type neighbors at $R_{\rm n} \leq0.5 r_{\rm vir,nei}$
  is examined with the K-S and MC tests.
The difference is confirmed with a significance level of 
  98 and $>$99\% from the K-S and MC tests, respectively, 
  in the local universe,
  and of 98$\%$ from both K-S and MC tests in high-$z$ universe.

To study the evolution of SSFRs depending on the morphology of
   the nearest neighbor,
   we plot, in Fig. \ref{fig-zevol}, the median SSFRs 
   of strongly-  and weakly-interacting galaxies 
   as a function of redshift
   in comparison with those for non-interacting galaxies.
We plot the SSFRs of interacting galaxies 
  normalized by those of non-interacting galaxies
  at each redshift bin
  in order to focus only on the relative ratio of SSFRs of interacting galaxies
  to those of non-interacting galaxies.
  
For the early-type neighbor case (left panel in Fig. \ref{fig-zevol}),
  the SSFRs of interacting galaxies are similar to
  or slightly smaller than those of non-interacting galaxies
  in all redshift bins.
However, for the late-type neighbor case (right panel),
  it is seen that the increased SSFRs 
  for weakly- and strongly-interacting galaxies
  are systematically larger than those for non-interacting galaxies 
  in all redshift bins.
The SSFRs of strongly- and weakly-interacting galaxies are, 
  on average, larger than 
  those of non-interacting galaxies by factors of 
  about 4.0$\pm$1.2 and 1.8$\pm$0.7, respectively.
If these increased SSFRs are simply due to the blending of two galaxies 
  with similar masses in one FIR beam without any enhanced SFA, 
  the SSFRs are expected to increase by a factor of only two.
In fact, the neighbor-separation scale of 
 $R_{\rm n} \sim 0.05 r_{\rm vir,nei}$
 ($\approx$15 $h^{-1}$ kpc for galaxies at $M_r=-21$)
 is important because the galaxies in pairs start to merge at this separation, 
 so the SFA starts to change abruptly \citep{pc09}.
Thus the increased SFR for strongly-interacting galaxies 
 ($R_{\rm n} \leq0.1 r_{\rm vir,nei}\approx$30 $h^{-1}$ kpc for $M_r=-21$)
 by a factor of four, really reflects the enhancement of 
 SFA due to a merger.

\subsection{Change in dust temperature as a function of pair separation}

In Fig. \ref{fig-tdust},
  we show the dependence of dust temperature on the distance to
  and the morphology of the nearest neighbor galaxy.
For local SDSS galaxies in (a),
  we plot the dust temperature derived from the flux density ratio ($S_{60}/S_{100}$)
  using Eq. (\ref{eq-tdust}) to increase the statistics.
For high-$z$ GOODS galaxies in (b),
  we do not use the dust temperature derived from the FIR flux density ratio
  because the large scatter in the correlation between
  {\it Herschel} flux density ratios and dust temperature
  makes statistics worse.

The dust temperature of both local and high-$z$ late-type galaxies
  appears to increase as they approach late-type neighbors,
  which is clearly seen at $R_{\rm n} \lesssim 0.1 r_{\rm vir,nei}$.
We also show the distribution of dust temperature in Fig. \ref{fig-hist}
  for interacting and non-interacting galaxies in SDSS and GOODS.
The MC test supports this different $T_{\rm dust}$ distribution between 
  strongly- and non-interacting galaxies having late-type neighbors
  with a significance level of 96 and $>$99\%
  for local and high-$z$ galaxies, respectively.
However, there are only two GOODS galaxies 
  at $R_{\rm n} \lesssim 0.1 r_{\rm vir,nei}$,
  so this trend needs to be checked with larger data set in future studies.
On the other hand,
  the dust temperature of local late-type galaxies with
  early-type neighbors at $R_{\rm n} \lesssim 0.1 r_{\rm vir,nei}$
  seems to be lower than or marginally similar to
  that of non-interacting galaxies.
Therefore, the difference in the dust temperature 
  of strongly-interacting galaxies with late- and early-type neighbors
  seems to be significant, 
  which is supported by the MC test with a significance level of 98\%.
For high-$z$ galaxies,
  there are few late-type galaxies having 
  close early-type neighbors at $R_{\rm n} \lesssim 0.1 r_{\rm vir,nei}$
  with $T_{\rm dust}$ measurements,
  so we can not study the trend.

\section{Discussion}\label{discuss}

The increase of SSFR for galaxies in pairs found in this study
  is consistent with results in previous studies
  at low redshifts (e.g., \citealt{ken87,bar00,lam03,nik04,woods07,hh11})
  and high redshifts (e.g., \citealt{lin07,wong11}).
However, note that the amount of increased SSFR compared to isolated galaxies
  is different depending on the studies because 
  the definition of interacting galaxies and 
  the observational selection effects are different.
The increase of dust temperature
  of local late-type galaxies strongly interacting
  with other late-type neighbors
  is also consistent with previous studies
  in the sense that the contribution of warm dust compared to cold dust 
  increases with the merging sequence \citep{tel88,xil04}.
However, we found the hint of this trend
  for high-$z$ galaxies at $0.4 \la z \la 1.2$ for the first time.
It is also interesting to see that the dust temperature of 
  local, late-type galaxies does not seem to increase
  when they are strongly interacting with early-type neighbors
  (see Fig. \ref{fig-tdust}).
Note also that the distance to the nearest neighbor 
    might not be a direct measure of the merging sequence because 
    galaxies in pairs would merge after several encounters
    and its orbital geometry is complicated.

We wish to emphasize that our analysis
  represents a major improvement compared to previous studies
  because we 
  1) have robust SFR measurements of galaxies with well-constrained FIR SEDs
   thanks to {\it Herschel} and {\it AKARI},
  2) have galaxy samples with high spectroscopic completeness 
   (i.e. $>85\%$ for SDSS galaxies at $m_r<17.77$ and 
   $71-86\%$ for GOODS galaxies at $m_z<23$),
  3) study only the galaxies that 
   are complete in terms of neighbor galaxies
   with stellar masses similar to target galaxies
   ($M_{r,\rm nei}\leq M_{r,\rm target}+0.5$),
  4) distinguish the morphology of galaxies in pairs,
   which is an important indicator of the gas fraction in galaxies, and
  5) use a distance to the nearest neighbor normalized by the virial radius
    of the neighbor instead of the physical distance
    in order to account for different masses and corresponding crossing time 
    of galaxy samples.

    
The dependence of SFR (or SSFR)
  on the morphology of the neighbor at $R_{\rm n}\approx 0.5 r_{\rm vir,nei}$ 
  seen in Fig. \ref{fig-neibl}
  may imply that the hydrodynamic interactions
  with the nearest neighbor play critical roles in triggering the SFA of galaxies
  in addition to the tidal interactions.
This dependence is also seen for local normal galaxies \citep{pc09,xu10}
  and local LIRGs and ULIRGs \citep{hwa10lirg},
  and can be explained as follows.  
If a late-type galaxy approaches a late-type neighbor 
  within half of the virial radius of the neighbor,
  the cold gas inflow into the central region of the target galaxy 
  from the neighbor galaxy as well as 
  from the disk of the target galaxy increases,
  which results in the enhanced SFA of the target galaxy.
Because of this starburst mode of compact star formation,
  the SSFR and the dust temperature are expected to
  increase as we observed (\citealt{elb11}; see also \citealt{cha07}).
Then when two galaxies finish merging,
  the end product of the merger will be bright due to the very recent SFA,
  and the new nearest neighbor galaxy of the merger product will be far away.
This may explain the existence of some non-interacting galaxies 
  at $R_{\rm n}>r_{\rm vir,nei}$
  with large SSFRs and $T_{\rm dust}$ as seen 
  in Figs. \ref{fig-neibl} and \ref{fig-tdust}.
  
On the other hand, if a galaxy approaches an early-type neighbor 
  within the virial radius of the neighbor,
  the hot gas of the early-type neighbor prevents 
  the galaxy from forming stars with cold gas, 
  and/or there is no inflow of cold gas from its early-type neighbor, 
  so the SFA of the galaxy is not boosted even if it has a close companion.
The SF quenching mechanisms of hot gas in early-type neighbors
  could be similar to those of a hot intracluster medium of galaxy clusters
  acting on late-type galaxies in it,
  which are hydrodynamic processes such as 
  thermal evaporation, strangulation, ram pressure stripping, or
  viscous stripping \citep{pc09,ph09}.
Indeed, the X-ray observations of galaxies in pairs 
  with mixed morphology
  show evidence for extended X-ray halos of the early type
  that surround the late type,
  which supports our interpretation \citep{gru07}.
In addition, the fact that 
  SSFRs in Fig. \ref{fig-zevol} are increased by similar factors
  in all redshift bins,
  suggests that similar physical mechanisms 
  (such as hydrodynamic interactions plus tidal interactions described above)
  affect on the SFA over at least 8 billion years.

The SFRs of galaxies are known to depend strongly on the local density,
  in the sense that spatially averaged SFRs of galaxies or 
  star-forming galaxy fractions 
  decrease as the background density increases in the local universe 
  (e.g., \citealt{lew02, gom03, park07, jhlee10env, hwa10lirg}).
In the high-$z$ universe,
  some studies found similar results (e.g., \citealt{pat09,fer10}), but
  there are also hints for an opposite trend 
  (i.e. increasing SFRs or star-forming galaxy 
  fractions with increasing the background density)
  known as the reversal of SFR-density relation 
  (e.g., \citealt{elb07, pop11, pop11mass}).
When we consider the morphology-density relation \citep{dre80},
  early(late)-type neighbors could be preferentially selected in high(low)-density regions,
  and the pair separation is correlated with the local density.
Therefore, the SFRs appear to depend on both large- and small-scale 
  (attributed to the nearest neighbor) environments.
One might expect that the difference in FIR properties depending on the morphology
  of and the distance to the nearest neighbor is simply due to the statistical correlation
  between the local density and the properties of nearest neighbor. 
Thus, it is necessary to disentangle the effects of both environments.
 
In previous studies,
  we have examined the effects of both large- and small-scale environments 
  on the SFA and
  the morphological transformation of SDSS galaxies 
  \citep{park08, pc09, hwa10lirg}, 
  and on the morphological transformation of GOODS galaxies \citep{hp09}. 
We found that the SFA of local galaxies is still strongly 
  affected by the nearest neighbors 
  differently depending on the morphology 
  {\it even at fixed large-scale background density} 
  (see Fig. 14 in \citealt{hwa10lirg} and Fig. 7 in \citealt{pc09}). 
The morphological transformation of high-$z$ galaxies shows a similar trend  
  (see Fig. 6 in \citealt{hp09}). 
Therefore, it is expected that the role of the large-scale environment 
  in the SFA for high-$z$ galaxies to be weak. 
We tried to investigate the change in SFRs of GOODS galaxies depending on 
  both large- and small-scale environments,
  and found that galaxies with early (late)-type neighbors
  are not preferentially selected in high (low)-density regions.
This implies that there is no bias in our results
  introduced by the difference in the large-scale environment
  (see also Fig. 6 in \citealt{hp09}).
However, we could not draw meaningful conclusions on
  the distinction of the effects of large- and small-scale environments on the SFA 
  due to the small number statistics, 
  which needs to be investigated with larger data set of 
  IR-detected galaxies in future studies.
On the other hand, a detailed analysis focusing on 
  the effects of large-scale environment on the SFA of high-$z$ galaxies 
  can be found in other studies based on similar GOODS data
 \citep{elb07,pop11,pop11mass}.

\section{Conclusions}\label{con}

Using the {\it Herschel}-selected galaxies in the GOODS fields
  and the {\it IRAS} plus {\it AKARI}-selected galaxies in the field of SDSS DR7,
  we studied the impact of galaxy-galaxy interactions
  on the FIR properties of galaxies and its evolution at $0<z<1.2$.
Our main results follow.
  
\begin{enumerate}
  
\item We find that the SFRs and SSFRs of galaxies, on average,
  depend on the morphology of and the distance to the nearest neighbor galaxy
  for all redshifts, within $0<z<1.2$.    
When a late-type galaxy has a close neighbor galaxy,
  the SFR and SSFR increase as it approaches a late-type neighbor,
  which is supported by K-S and MC tests  with a significance level of $>$99\%.
However, the SFR and SSFR
  decrease or do not change much as it approaches an early-type neighbor.
The bifurcations of SFRs and SSFRs depending on the neighbor's morphology
  are seen at $R_n\approx 0.5 r_{\rm vir,nei}$,
  which is also supported by K-S and MC tests with a significance level of $>$98\%. 
  
\item For the redshift range $0<z<1.2$,
  the SSFRs of late-type galaxies having late-type neighbors
  are increased by factors of about 1.8$\pm$0.7 and 4.0$\pm$1.2, respectively,
  for the cases of 
  weakly- ($R_{\rm n} \leq0.5 r_{\rm vir,nei}$)
  and strongly-interacting ($R_{\rm n} \leq0.1 r_{\rm vir,nei}$) galaxies
  compared to those of non-interacting galaxies.

\item The dust temperature
  of both local and high-$z$ late-type galaxies 
  strongly interacting with late-type neighbors
  appears to be higher than that of non-interacting galaxies
  with a significance leve of $96-99$\%.
However,   
  the dust temperature
  of local late-type galaxies strongly interacting with early-type neighbors
  seems to be lower than or similar to that of non-interacting galaxies.
  
\end{enumerate}

Our results suggest that galaxy-galaxy interactions and mergers
  have been strongly affecting the SFA and the dust properties of star-forming galaxies
  over at least 8 billion years.

\begin{acknowledgements}
We would like to thank the anonymous referee for constructive comments that 
  helped us to improve the manuscript.
HSH and DE acknowledge the support of the Centre National d'Etudes Spatiales (CNES).
DE acknowledges financial support from the French Agence Nationale de la Recherche (ANR) project ``HUGE'', 
ANR-09-BLAN-0224.
VC would like to acknowledge partial support from the EU ToK grant 39965 and FP7-REGPOT 206469.
PACS has been developed by a consortium of institutes led by MPE (Germany) and including 
UVIE (Austria); KU Leuven, CSL, IMEC (Belgium); CEA, LAM (France); MPIA (Germany);
INAFIFSI/OAA/OAP/OAT, LENS, SISSA (Italy); IAC (Spain). This development has been supported by 
the funding agencies BMVIT (Austria), ESA-PRODEX (Belgium), CEA/CNES (France), DLR (Germany),
ASI/INAF (Italy), and CICYT/MCYT (Spain). 
SPIRE has been developed by a consortium of institutes led by Cardiff University (UK) and 
including Univ. Lethbridge (Canada); NAOC (China); CEA, LAM (France); IFSI, Univ. Padua (Italy); 
IAC (Spain); SNSB (Sweden); Imperial College London, RAL, UCL-MSSL, UKATC, 
Univ. Sussex (UK); and Caltech, JPL, NHSC, Univ. Colorado (USA). This development has been 
supported by national funding agencies: CSA (Canada); NAOC (China); CEA, CNES, CNRS (France); 
ASI (Italy); MCINN (Spain); Stockholm Observatory (Sweden); STFC (UK); and NASA (USA).
This research is based on observations with AKARI, a JAXA project with the participation of ESA.
Funding for the SDSS and SDSS-II has been provided by the Alfred P. Sloan 
Foundation, the Participating Institutions, the National Science 
Foundation, the U.S. Department of Energy, the National Aeronautics and 
Space Administration, the Japanese Monbukagakusho, the Max Planck 
Society, and the Higher Education Funding Council for England. 
The SDSS Web Site is http://www.sdss.org/.
The SDSS is managed by the Astrophysical Research Consortium for the 
Participating Institutions. The Participating Institutions are the 
American Museum of Natural History, Astrophysical Institute Potsdam, 
University of Basel, Cambridge University, Case Western Reserve University, 
University of Chicago, Drexel University, Fermilab, the Institute for 
Advanced Study, the Japan Participation Group, Johns Hopkins University, 
the Joint Institute for Nuclear Astrophysics, the Kavli Institute for 
Particle Astrophysics and Cosmology, the Korean Scientist Group, the 
Chinese Academy of Sciences (LAMOST), Los Alamos National Laboratory, 
the Max-Planck-Institute for Astronomy (MPIA), the Max-Planck-Institute 
for Astrophysics (MPA), New Mexico State University, Ohio State University, 
University of Pittsburgh, University of Portsmouth, Princeton University,
the United States Naval Observatory, and the University of Washington. 
This research has made use of the NASA/IPAC Extragalactic Database (NED) 
which is operated by the Jet Propulsion Laboratory, California Institute of Technology, 
under contract with the National Aeronautics and Space Administration.
\end{acknowledgements}

\bibliographystyle{aa} 
\bibliography{ref_hshwang} 

\begin{thebibliography}{115}
\expandafter\ifx\csname natexlab\endcsname\relax\def\natexlab#1{#1}\fi

\bibitem[{{Abazajian} {et~al.}(2009){Abazajian}, {Adelman-McCarthy},
  {Ag{\"u}eros}, {Allam}, {Allende Prieto}, {An}, {Anderson}, {Anderson},
  {Annis}, {Bahcall}, {Bailer-Jones}, {Barentine}, {Bassett}, {Becker},
  {Beers}, {Bell}, {Belokurov}, {Berlind}, {Berman}, {Bernardi}, {Bickerton},
  {Bizyaev}, {Blakeslee}, {Blanton}, {Bochanski}, {Boroski}, {Brewington},
  {Brinchmann}, {Brinkmann}, {Brunner}, {Budav{\'a}ri}, {Carey}, {Carliles},
  {Carr}, {Castander}, {Cinabro}, {Connolly}, {Csabai}, {Cunha}, {Czarapata},
  {Davenport}, {de Haas}, {Dilday}, {Doi}, {Eisenstein}, {Evans}, {Evans},
  {Fan}, {Friedman}, {Frieman}, {Fukugita}, {G{\"a}nsicke}, {Gates},
  {Gillespie}, {Gilmore}, {Gonzalez}, {Gonzalez}, {Grebel}, {Gunn},
  {Gy{\"o}ry}, {Hall}, {Harding}, {Harris}, {Harvanek}, {Hawley}, {Hayes},
  {Heckman}, {Hendry}, {Hennessy}, {Hindsley}, {Hoblitt}, {Hogan}, {Hogg},
  {Holtzman}, {Hyde}, {Ichikawa}, {Ichikawa}, {Im}, {Ivezi{\'c}}, {Jester},
  {Jiang}, {Johnson}, {Jorgensen}, {Juri{\'c}}, {Kent}, {Kessler}, {Kleinman},
  {Knapp}, {Konishi}, {Kron}, {Krzesinski}, {Kuropatkin}, {Lampeitl},
  {Lebedeva}, {Lee}, {Lee}, {Leger}, {L{\'e}pine}, {Li}, {Lima}, {Lin}, {Long},
  {Loomis}, {Loveday}, {Lupton}, {Magnier}, {Malanushenko}, {Malanushenko},
  {Mandelbaum}, {Margon}, {Marriner}, {Mart{\'{\i}}nez-Delgado}, {Matsubara},
  {McGehee}, {McKay}, {Meiksin}, {Morrison}, {Mullally}, {Munn}, {Murphy},
  {Nash}, {Nebot}, {Neilsen}, {Newberg}, {Newman}, {Nichol}, {Nicinski},
  {Nieto-Santisteban}, {Nitta}, {Okamura}, {Oravetz}, {Ostriker}, {Owen},
  {Padmanabhan}, {Pan}, {Park}, {Pauls}, {Peoples}, {Percival}, {Pier}, {Pope},
  {Pourbaix}, {Price}, {Purger}, {Quinn}, {Raddick}, {Fiorentin}, {Richards},
  {Richmond}, {Riess}, {Rix}, {Rockosi}, {Sako}, {Schlegel}, {Schneider},
  {Scholz}, {Schreiber}, {Schwope}, {Seljak}, {Sesar}, {Sheldon}, {Shimasaku},
  {Sibley}, {Simmons}, {Sivarani}, {Smith}, {Smith}, {Smol{\v c}i{\'c}},
  {Snedden}, {Stebbins}, {Steinmetz}, {Stoughton}, {Strauss}, {Subba Rao},
  {Suto}, {Szalay}, {Szapudi}, {Szkody}, {Tanaka}, {Tegmark}, {Teodoro},
  {Thakar}, {Tremonti}, {Tucker}, {Uomoto}, {Vanden Berk}, {Vandenberg},
  {Vidrih}, {Vogeley}, {Voges}, {Vogt}, {Wadadekar}, {Watters}, {Weinberg},
  {West}, {White}, {Wilhite}, {Wonders}, {Yanny}, {Yocum}, {York}, {Zehavi},
  {Zibetti}, \& {Zucker}}]{aba09}
{Abazajian}, K.~N., {Adelman-McCarthy}, J.~K., {Ag{\"u}eros}, M.~A., {et~al.}
  2009, \apjs, 182, 543

\bibitem[{{Alonso} {et~al.}(2004){Alonso}, {Tissera}, {Coldwell}, \&
  {Lambas}}]{alo04}
{Alonso}, M.~S., {Tissera}, P.~B., {Coldwell}, G., \& {Lambas}, D.~G. 2004,
  \mnras, 352, 1081

\bibitem[{{Ann} {et~al.}(2008){Ann}, {Park}, \& {Choi}}]{ann08}
{Ann}, H.~B., {Park}, C., \& {Choi}, Y. 2008, \mnras, 389, 86

\bibitem[{{Balestra} {et~al.}(2010){Balestra}, {Mainieri}, {Popesso},
  {Dickinson}, {Nonino}, {Rosati}, {Teimoorinia}, {Vanzella}, {Cristiani},
  {Cesarsky}, {Fosbury}, {Kuntschner}, \& {Rettura}}]{bal10}
{Balestra}, I., {Mainieri}, V., {Popesso}, P., {et~al.} 2010, \aap, 512, A12

\bibitem[{{Barger} {et~al.}(2008){Barger}, {Cowie}, \& {Wang}}]{bar08}
{Barger}, A.~J., {Cowie}, L.~L., \& {Wang}, W.-H. 2008, \apj, 689, 687

\bibitem[{{Barton} {et~al.}(2000){Barton}, {Geller}, \& {Kenyon}}]{bar00}
{Barton}, E.~J., {Geller}, M.~J., \& {Kenyon}, S.~J. 2000, \apj, 530, 660

\bibitem[{{Bergvall} {et~al.}(2003){Bergvall}, {Laurikainen}, \&
  {Aalto}}]{ber03}
{Bergvall}, N., {Laurikainen}, E., \& {Aalto}, S. 2003, \aap, 405, 31

\bibitem[{{Blanton} \& {Roweis}(2007)}]{bla07kcorr}
{Blanton}, M.~R. \& {Roweis}, S. 2007, \aj, 133, 734

\bibitem[{{Bridge} {et~al.}(2010){Bridge}, {Carlberg}, \& {Sullivan}}]{bri10}
{Bridge}, C.~R., {Carlberg}, R.~G., \& {Sullivan}, M. 2010, \apj, 709, 1067

\bibitem[{{Brinchmann} {et~al.}(2004){Brinchmann}, {Charlot}, {White},
  {Tremonti}, {Kauffmann}, {Heckman}, \& {Brinkmann}}]{bri04}
{Brinchmann}, J., {Charlot}, S., {White}, S.~D.~M., {et~al.} 2004, \mnras, 351,
  1151

\bibitem[{{Chanial} {et~al.}(2007){Chanial}, {Flores}, {Guiderdoni}, {Elbaz},
  {Hammer}, \& {Vigroux}}]{cha07}
{Chanial}, P., {Flores}, H., {Guiderdoni}, B., {et~al.} 2007, \aap, 462, 81

\bibitem[{{Chary} \& {Elbaz}(2001)}]{ce01}
{Chary}, R. \& {Elbaz}, D. 2001, \apj, 556, 562

\bibitem[{{Choi} {et~al.}(2010){Choi}, {Han}, \& {Kim}}]{choi10}
{Choi}, Y., {Han}, D., \& {Kim}, S.~S. 2010, Journal of Korean Astronomical
  Society, 43, 191

\bibitem[{{Choi} {et~al.}(2007){Choi}, {Park}, \& {Vogeley}}]{choi07}
{Choi}, Y., {Park}, C., \& {Vogeley}, M.~S. 2007, \apj, 658, 884

\bibitem[{{Cohen} {et~al.}(2000){Cohen}, {Hogg}, {Blandford}, {Cowie}, {Hu},
  {Songaila}, {Shopbell}, \& {Richberg}}]{coh00}
{Cohen}, J.~G., {Hogg}, D.~W., {Blandford}, R., {et~al.} 2000, \apj, 538, 29

\bibitem[{{Condon} {et~al.}(1982){Condon}, {Condon}, {Gisler}, \&
  {Puschell}}]{con82}
{Condon}, J.~J., {Condon}, M.~A., {Gisler}, G., \& {Puschell}, J.~J. 1982,
  \apj, 252, 102

\bibitem[{{Cooper} {et~al.}(2011){Cooper}, {Aird}, {Coil}, {Davis}, {Faber},
  {Juneau}, {Lotz}, {Nandra}, {Newman}, {Willmer}, \& {Yan}}]{coo11}
{Cooper}, M.~C., {Aird}, J.~A., {Coil}, A.~L., {et~al.} 2011, \apjs, 193, 14

\bibitem[{{Cowie} {et~al.}(2004){Cowie}, {Barger}, {Hu}, {Capak}, \&
  {Songaila}}]{cow04}
{Cowie}, L.~L., {Barger}, A.~J., {Hu}, E.~M., {Capak}, P., \& {Songaila}, A.
  2004, \aj, 127, 3137

\bibitem[{{Daddi} {et~al.}(2007){Daddi}, {Dickinson}, {Morrison}, {Chary},
  {Cimatti}, {Elbaz}, {Frayer}, {Renzini}, {Pope}, {Alexander}, {Bauer},
  {Giavalisco}, {Huynh}, {Kurk}, \& {Mignoli}}]{dad07}
{Daddi}, E., {Dickinson}, M., {Morrison}, G., {et~al.} 2007, \apj, 670, 156

\bibitem[{{Darg} {et~al.}(2010){Darg}, {Kaviraj}, {Lintott}, {Schawinski},
  {Sarzi}, {Bamford}, {Silk}, {Andreescu}, {Murray}, {Nichol}, {Raddick},
  {Slosar}, {Szalay}, {Thomas}, \& {Vandenberg}}]{darg10}
{Darg}, D.~W., {Kaviraj}, S., {Lintott}, C.~J., {et~al.} 2010, \mnras, 401,
  1552

\bibitem[{{De Propris} {et~al.}(2005){De Propris}, {Liske}, {Driver}, {Allen},
  \& {Cross}}]{deprop05}
{De Propris}, R., {Liske}, J., {Driver}, S.~P., {Allen}, P.~D., \& {Cross},
  N.~J.~G. 2005, \aj, 130, 1516

\bibitem[{{de Ravel} {et~al.}(2009){de Ravel}, {Le F{\`e}vre}, {Tresse},
  {Bottini}, {Garilli}, {Le Brun}, {Maccagni}, {Scaramella}, {Scodeggio},
  {Vettolani}, {Zanichelli}, {Adami}, {Arnouts}, {Bardelli}, {Bolzonella},
  {Cappi}, {Charlot}, {Ciliegi}, {Contini}, {Foucaud}, {Franzetti},
  {Gavignaud}, {Guzzo}, {Ilbert}, {Iovino}, {Lamareille}, {McCracken},
  {Marano}, {Marinoni}, {Mazure}, {Meneux}, {Merighi}, {Paltani}, {Pell{\`o}},
  {Pollo}, {Pozzetti}, {Radovich}, {Vergani}, {Zamorani}, {Zucca}, {Bondi},
  {Bongiorno}, {Brinchmann}, {Cucciati}, {de La Torre}, {Gregorini}, {Memeo},
  {Perez-Montero}, {Mellier}, {Merluzzi}, \& {Temporin}}]{der09}
{de Ravel}, L., {Le F{\`e}vre}, O., {Tresse}, L., {et~al.} 2009, \aap, 498, 379

\bibitem[{{Dickinson} {et~al.}(2003){Dickinson}, {Giavalisco}, \& {GOODS
  Team}}]{dic03}
{Dickinson}, M., {Giavalisco}, M., \& {GOODS Team}. 2003, in The Mass of
  Galaxies at Low and High Redshift, ed. {R.~Bender \& A.~Renzini}, 324--+

\bibitem[{{Dressler}(1980)}]{dre80}
{Dressler}, A. 1980, \apj, 236, 351

\bibitem[{{Elbaz} {et~al.}(2007){Elbaz}, {Daddi}, {Le Borgne}, {Dickinson},
  {Alexander}, {Chary}, {Starck}, {Brandt}, {Kitzbichler}, {MacDonald},
  {Nonino}, {Popesso}, {Stern}, \& {Vanzella}}]{elb07}
{Elbaz}, D., {Daddi}, E., {Le Borgne}, D., {et~al.} 2007, \aap, 468, 33

\bibitem[{{Elbaz} {et~al.}(2010){Elbaz}, {Hwang}, {Magnelli}, {Daddi},
  {Aussel}, {Altieri}, {Amblard}, {Andreani}, {Arumugam}, {Auld}, {Babbedge},
  {Berta}, {Blain}, {Bock}, {Bongiovanni}, {Boselli}, {Buat}, {Burgarella},
  {Castro-Rodriguez}, {Cava}, {Cepa}, {Chanial}, {Chary}, {Cimatti},
  {Clements}, {Conley}, {Conversi}, {Cooray}, {Dickinson}, {Dominguez},
  {Dowell}, {Dunlop}, {Dwek}, {Eales}, {Farrah}, {F{\"o}rster Schreiber},
  {Fox}, {Franceschini}, {Gear}, {Genzel}, {Glenn}, {Griffin}, {Gruppioni},
  {Halpern}, {Hatziminaoglou}, {Ibar}, {Isaak}, {Ivison}, {Lagache}, {Le
  Borgne}, {Le Floc'h}, {Levenson}, {Lu}, {Lutz}, {Madden}, {Maffei}, {Magdis},
  {Mainetti}, {Maiolino}, {Marchetti}, {Mortier}, {Nguyen}, {Nordon},
  {O'Halloran}, {Okumura}, {Oliver}, {Omont}, {Page}, {Panuzzo},
  {Papageorgiou}, {Pearson}, {Perez Fournon}, {P{\'e}rez Garc{\'{\i}}a},
  {Poglitsch}, {Pohlen}, {Popesso}, {Pozzi}, {Rawlings}, {Rigopoulou},
  {Riguccini}, {Rizzo}, {Rodighiero}, {Roseboom}, {Rowan-Robinson},
  {Saintonge}, {Sanchez Portal}, {Santini}, {Sauvage}, {Schulz}, {Scott},
  {Seymour}, {Shao}, {Shupe}, {Smith}, {Stevens}, {Sturm}, {Symeonidis},
  {Tacconi}, {Trichas}, {Tugwell}, {Vaccari}, {Valtchanov}, {Vieira},
  {Vigroux}, {Wang}, {Ward}, {Wright}, {Xu}, \& {Zemcov}}]{elb10}
{Elbaz}, D., {Hwang}, H.~S., {Magnelli}, B., {et~al.} 2010, \aap, 518, L29

\bibitem[{{Elbaz} {et~al.}(2011)}]{elb11}
{Elbaz}, D. {et~al.} 2011, \aap, in press (arXiv:1105.2537)

\bibitem[{{Ellison} {et~al.}(2008){Ellison}, {Patton}, {Simard}, \&
  {McConnachie}}]{ell08}
{Ellison}, S.~L., {Patton}, D.~R., {Simard}, L., \& {McConnachie}, A.~W. 2008,
  \aj, 135, 1877

\bibitem[{{Feruglio} {et~al.}(2010){Feruglio}, {Aussel}, {Le Floc'h}, {Ilbert},
  {Salvato}, {Capak}, {Fiore}, {Kartaltepe}, {Sanders}, {Scoville},
  {Koekemoer}, \& {Ideue}}]{fer10}
{Feruglio}, C., {Aussel}, H., {Le Floc'h}, E., {et~al.} 2010, \apj, 721, 607

\bibitem[{{Fioc} \& {Rocca-Volmerange}(1999)}]{fr99}
{Fioc}, M. \& {Rocca-Volmerange}, B. 1999, astro-ph/9912179

\bibitem[{{Gallazzi} {et~al.}(2005){Gallazzi}, {Charlot}, {Brinchmann},
  {White}, \& {Tremonti}}]{gal05}
{Gallazzi}, A., {Charlot}, S., {Brinchmann}, J., {White}, S.~D.~M., \&
  {Tremonti}, C.~A. 2005, \mnras, 362, 41

\bibitem[{{Geller} {et~al.}(2006){Geller}, {Kenyon}, {Barton}, {Jarrett}, \&
  {Kewley}}]{gel06}
{Geller}, M.~J., {Kenyon}, S.~J., {Barton}, E.~J., {Jarrett}, T.~H., \&
  {Kewley}, L.~J. 2006, \aj, 132, 2243

\bibitem[{{Giavalisco} {et~al.}(2004){Giavalisco}, {Ferguson}, {Koekemoer},
  {Dickinson}, {Alexander}, {Bauer}, {Bergeron}, {Biagetti}, {Brandt},
  {Casertano}, {Cesarsky}, {Chatzichristou}, {Conselice}, {Cristiani}, {Da
  Costa}, {Dahlen}, {de Mello}, {Eisenhardt}, {Erben}, {Fall}, {Fassnacht},
  {Fosbury}, {Fruchter}, {Gardner}, {Grogin}, {Hook}, {Hornschemeier}, {Idzi},
  {Jogee}, {Kretchmer}, {Laidler}, {Lee}, {Livio}, {Lucas}, {Madau},
  {Mobasher}, {Moustakas}, {Nonino}, {Padovani}, {Papovich}, {Park},
  {Ravindranath}, {Renzini}, {Richardson}, {Riess}, {Rosati}, {Schirmer},
  {Schreier}, {Somerville}, {Spinrad}, {Stern}, {Stiavelli}, {Strolger},
  {Urry}, {Vandame}, {Williams}, \& {Wolf}}]{gia04}
{Giavalisco}, M., {Ferguson}, H.~C., {Koekemoer}, A.~M., {et~al.} 2004, \apjl,
  600, L93

\bibitem[{{G{\'o}mez} {et~al.}(2003){G{\'o}mez}, {Nichol}, {Miller}, {Balogh},
  {Goto}, {Zabludoff}, {Romer}, {Bernardi}, {Sheth}, {Hopkins}, {Castander},
  {Connolly}, {Schneider}, {Brinkmann}, {Lamb}, {SubbaRao}, \& {York}}]{gom03}
{G{\'o}mez}, P.~L., {Nichol}, R.~C., {Miller}, C.~J., {et~al.} 2003, \apj, 584,
  210

\bibitem[{{Griffin} {et~al.}(2010){Griffin}, {Abergel}, {Abreu}, {Ade},
  {Andr{\'e}}, {Augueres}, {Babbedge}, {Bae}, {Baillie}, {Baluteau}, {Barlow},
  {Bendo}, {Benielli}, {Bock}, {Bonhomme}, {Brisbin}, {Brockley-Blatt},
  {Caldwell}, {Cara}, {Castro-Rodriguez}, {Cerulli}, {Chanial}, {Chen},
  {Clark}, {Clements}, {Clerc}, {Coker}, {Communal}, {Conversi}, {Cox},
  {Crumb}, {Cunningham}, {Daly}, {Davis}, {de Antoni}, {Delderfield}, {Devin},
  {di Giorgio}, {Didschuns}, {Dohlen}, {Donati}, {Dowell}, {Dowell}, {Duband},
  {Dumaye}, {Emery}, {Ferlet}, {Ferrand}, {Fontignie}, {Fox}, {Franceschini},
  {Frerking}, {Fulton}, {Garcia}, {Gastaud}, {Gear}, {Glenn}, {Goizel},
  {Griffin}, {Grundy}, {Guest}, {Guillemet}, {Hargrave}, {Harwit}, {Hastings},
  {Hatziminaoglou}, {Herman}, {Hinde}, {Hristov}, {Huang}, {Imhof}, {Isaak},
  {Israelsson}, {Ivison}, {Jennings}, {Kiernan}, {King}, {Lange}, {Latter},
  {Laurent}, {Laurent}, {Leeks}, {Lellouch}, {Levenson}, {Li}, {Li},
  {Lilienthal}, {Lim}, {Liu}, {Lu}, {Madden}, {Mainetti}, {Marliani}, {McKay},
  {Mercier}, {Molinari}, {Morris}, {Moseley}, {Mulder}, {Mur}, {Naylor},
  {Nguyen}, {O'Halloran}, {Oliver}, {Olofsson}, {Olofsson}, {Orfei}, {Page},
  {Pain}, {Panuzzo}, {Papageorgiou}, {Parks}, {Parr-Burman}, {Pearce},
  {Pearson}, {P{\'e}rez-Fournon}, {Pinsard}, {Pisano}, {Podosek}, {Pohlen},
  {Polehampton}, {Pouliquen}, {Rigopoulou}, {Rizzo}, {Roseboom}, {Roussel},
  {Rowan-Robinson}, {Rownd}, {Saraceno}, {Sauvage}, {Savage}, {Savini},
  {Sawyer}, {Scharmberg}, {Schmitt}, {Schneider}, {Schulz}, {Schwartz},
  {Shafer}, {Shupe}, {Sibthorpe}, {Sidher}, {Smith}, {Smith}, {Smith},
  {Spencer}, {Stobie}, {Sudiwala}, {Sukhatme}, {Surace}, {Stevens}, {Swinyard},
  {Trichas}, {Tourette}, {Triou}, {Tseng}, {Tucker}, {Turner}, {Vaccari},
  {Valtchanov}, {Vigroux}, {Virique}, {Voellmer}, {Walker}, {Ward}, {Waskett},
  {Weilert}, {Wesson}, {White}, {Whitehouse}, {Wilson}, {Winter}, {Woodcraft},
  {Wright}, {Xu}, {Zavagno}, {Zemcov}, {Zhang}, \& {Zonca}}]{gri10}
{Griffin}, M.~J., {Abergel}, A., {Abreu}, A., {et~al.} 2010, \aap, 518, L3

\bibitem[{{Gr{\"u}tzbauch} {et~al.}(2007){Gr{\"u}tzbauch}, {Trinchieri},
  {Rampazzo}, {Held}, {Rizzi}, {Sulentic}, \& {Zeilinger}}]{gru07}
{Gr{\"u}tzbauch}, R., {Trinchieri}, G., {Rampazzo}, R., {et~al.} 2007, \aj,
  133, 220

\bibitem[{{Hern{\'a}ndez-Toledo} {et~al.}(2005){Hern{\'a}ndez-Toledo},
  {Avila-Reese}, {Conselice}, \& {Puerari}}]{her05}
{Hern{\'a}ndez-Toledo}, H.~M., {Avila-Reese}, V., {Conselice}, C.~J., \&
  {Puerari}, I. 2005, \aj, 129, 682

\bibitem[{{Huang} \& {Hwang}(2011)}]{hh11}
{Huang}, M.-L. \& {Hwang}, C.-Y. 2011, \apj, 734, 99

\bibitem[{{Hwang} {et~al.}(2010{\natexlab{a}}){Hwang}, {Elbaz}, {Lee}, {Jeong},
  {Park}, {Lee}, \& {Lee}}]{hwa10lirg}
{Hwang}, H.~S., {Elbaz}, D., {Lee}, J.~C., {et~al.} 2010{\natexlab{a}}, \aap,
  522, A33

\bibitem[{{Hwang} {et~al.}(2010{\natexlab{b}}){Hwang}, {Elbaz}, {Magdis},
  {Daddi}, {Symeonidis}, {Altieri}, {Amblard}, {Andreani}, {Arumugam}, {Auld},
  {Aussel}, {Babbedge}, {Berta}, {Blain}, {Bock}, {Bongiovanni}, {Boselli},
  {Buat}, {Burgarella}, {Castro-Rodr{\'{\i}}guez}, {Cava}, {Cepa}, {Chanial},
  {Chapin}, {Chary}, {Cimatti}, {Clements}, {Conley}, {Conversi}, {Cooray},
  {Dannerbauer}, {Dickinson}, {Dominguez}, {Dowell}, {Dunlop}, {Dwek}, {Eales},
  {Farrah}, {Schreiber}, {Fox}, {Franceschini}, {Gear}, {Genzel}, {Glenn},
  {Griffin}, {Gruppioni}, {Halpern}, {Hatziminaoglou}, {Ibar}, {Isaak},
  {Ivison}, {Jeong}, {Lagache}, {Le Borgne}, {Le Floc'h}, {Lee}, {Lee}, {Lee},
  {Levenson}, {Lu}, {Lutz}, {Madden}, {Maffei}, {Magnelli}, {Mainetti},
  {Maiolino}, {Marchetti}, {Mortier}, {Nguyen}, {Nordon}, {O'Halloran},
  {Okumura}, {Oliver}, {Omont}, {Page}, {Panuzzo}, {Papageorgiou}, {Pearson},
  {P{\'e}rez-Fournon}, {Garc{\'{\i}}a}, {Poglitsch}, {Pohlen}, {Popesso},
  {Pozzi}, {Rawlings}, {Rigopoulou}, {Riguccini}, {Rizzo}, {Rodighiero},
  {Roseboom}, {Rowan-Robinson}, {Saintonge}, {Portal}, {Santini}, {Sauvage},
  {Schulz}, {Scott}, {Seymour}, {Shao}, {Shupe}, {Smith}, {Stevens}, {Sturm},
  {Tacconi}, {Trichas}, {Tugwell}, {Vaccari}, {Valtchanov}, {Vieira},
  {Vigroux}, {Wang}, {Ward}, {Wright}, {Xu}, \& {Zemcov}}]{hwa10tdust}
{Hwang}, H.~S., {Elbaz}, D., {Magdis}, G., {et~al.} 2010{\natexlab{b}}, \mnras,
  409, 75

\bibitem[{{Hwang} \& {Park}(2009)}]{hp09}
{Hwang}, H.~S. \& {Park}, C. 2009, \apj, 700, 791

\bibitem[{{Isobe} {et~al.}(1990){Isobe}, {Feigelson}, {Akritas}, \&
  {Babu}}]{iso90}
{Isobe}, T., {Feigelson}, E.~D., {Akritas}, M.~G., \& {Babu}, G.~J. 1990, \apj,
  364, 104

\bibitem[{{Kartaltepe} {et~al.}(2007){Kartaltepe}, {Sanders}, {Scoville},
  {Calzetti}, {Capak}, {Koekemoer}, {Mobasher}, {Murayama}, {Salvato},
  {Sasaki}, \& {Taniguchi}}]{kar07}
{Kartaltepe}, J.~S., {Sanders}, D.~B., {Scoville}, N.~Z., {et~al.} 2007, \apjs,
  172, 320

\bibitem[{{Kauffmann} {et~al.}(2003){Kauffmann}, {Heckman}, {White}, {Charlot},
  {Tremonti}, {Brinchmann}, {Bruzual}, {Peng}, {Seibert}, {Bernardi},
  {Blanton}, {Brinkmann}, {Castander}, {Cs{\'a}bai}, {Fukugita}, {Ivezic},
  {Munn}, {Nichol}, {Padmanabhan}, {Thakar}, {Weinberg}, \& {York}}]{kau03}
{Kauffmann}, G., {Heckman}, T.~M., {White}, S.~D.~M., {et~al.} 2003, \mnras,
  341, 33

\bibitem[{{Kawada} {et~al.}(2007){Kawada}, {Baba}, {Barthel}, {Clements},
  {Cohen}, {Doi}, {Figueredo}, {Fujiwara}, {Goto}, {Hasegawa}, {Hibi}, {Hirao},
  {Hiromoto}, {Jeong}, {Kaneda}, {Kawai}, {Kawamura}, {Kester}, {Kii},
  {Kobayashi}, {Kwon}, {Lee}, {Makiuti}, {Matsuo}, {Matsuura}, {M{\"u}ller},
  {Murakami}, {Nagata}, {Nakagawa}, {Narita}, {Noda}, {Oh}, {Okada}, {Okuda},
  {Oliver}, {Ootsubo}, {Pak}, {Park}, {Pearson}, {Rowan-Robinson}, {Saito},
  {Salama}, {Sato}, {Savage}, {Serjeant}, {Shibai}, {Shirahata}, {Sohn},
  {Suzuki}, {Takagi}, {Takahashi}, {Thomson}, {Usui}, {Verdugo}, {Watabe},
  {White}, {Wang}, {Yamamura}, {Yamauchi}, \& {Yasuda}}]{kaw07}
{Kawada}, M., {Baba}, H., {Barthel}, P.~D., {et~al.} 2007, \pasj, 59, 389

\bibitem[{{Keel} {et~al.}(1985){Keel}, {Kennicutt}, {Hummel}, \& {van der
  Hulst}}]{keel85}
{Keel}, W.~C., {Kennicutt}, Jr., R.~C., {Hummel}, E., \& {van der Hulst}, J.~M.
  1985, \aj, 90, 708

\bibitem[{{Kennicutt}(1998)}]{ken98}
{Kennicutt}, Jr., R.~C. 1998, \araa, 36, 189

\bibitem[{{Kennicutt} {et~al.}(1987){Kennicutt}, {Roettiger}, {Keel}, {van der
  Hulst}, \& {Hummel}}]{ken87}
{Kennicutt}, Jr., R.~C., {Roettiger}, K.~A., {Keel}, W.~C., {van der Hulst},
  J.~M., \& {Hummel}, E. 1987, \aj, 93, 1011

\bibitem[{{Klaas} {et~al.}(2001){Klaas}, {Haas}, {M{\"u}ller}, {Chini},
  {Schulz}, {Coulson}, {Hippelein}, {Wilke}, {Albrecht}, \& {Lemke}}]{klaas01}
{Klaas}, U., {Haas}, M., {M{\"u}ller}, S.~A.~H., {et~al.} 2001, \aap, 379, 823

\bibitem[{{Knapen} \& {James}(2009)}]{kna09}
{Knapen}, J.~H. \& {James}, P.~A. 2009, \apj, 698, 1437

\bibitem[{{Kroupa}(2001)}]{kro01}
{Kroupa}, P. 2001, \mnras, 322, 231

\bibitem[{{Kurk} {et~al.}(2009){Kurk}, {Cimatti}, {Zamorani}, {Halliday},
  {Mignoli}, {Pozzetti}, {Daddi}, {Rosati}, {Dickinson}, {Bolzonella},
  {Cassata}, {Renzini}, {Franceschini}, {Rodighiero}, \& {Berta}}]{kurk09}
{Kurk}, J., {Cimatti}, A., {Zamorani}, G., {et~al.} 2009, \aap, 504, 331

\bibitem[{{Lambas} {et~al.}(2003){Lambas}, {Tissera}, {Alonso}, \&
  {Coldwell}}]{lam03}
{Lambas}, D.~G., {Tissera}, P.~B., {Alonso}, M.~S., \& {Coldwell}, G. 2003,
  \mnras, 346, 1189

\bibitem[{{Larson} \& {Tinsley}(1978)}]{lt78}
{Larson}, R.~B. \& {Tinsley}, B.~M. 1978, \apj, 219, 46

\bibitem[{{Le Borgne} \& {Rocca-Volmerange}(2002)}]{lr02}
{Le Borgne}, D. \& {Rocca-Volmerange}, B. 2002, \aap, 386, 446

\bibitem[{{Le F{\`e}vre} {et~al.}(2004){Le F{\`e}vre}, {Vettolani}, {Paltani},
  {Tresse}, {Zamorani}, {Le Brun}, {Moreau}, {Bottini}, {Maccagni}, {Picat},
  {Scaramella}, {Scodeggio}, {Zanichelli}, {Adami}, {Arnouts}, {Bardelli},
  {Bolzonella}, {Cappi}, {Charlot}, {Contini}, {Foucaud}, {Franzetti},
  {Garilli}, {Gavignaud}, {Guzzo}, {Ilbert}, {Iovino}, {McCracken}, {Mancini},
  {Marano}, {Marinoni}, {Mathez}, {Mazure}, {Meneux}, {Merighi}, {Pell{\`o}},
  {Pollo}, {Pozzetti}, {Radovich}, {Zucca}, {Arnaboldi}, {Bondi}, {Bongiorno},
  {Busarello}, {Ciliegi}, {Gregorini}, {Mellier}, {Merluzzi}, {Ripepi}, \&
  {Rizzo}}]{lef04}
{Le F{\`e}vre}, O., {Vettolani}, G., {Paltani}, S., {et~al.} 2004, \aap, 428,
  1043

\bibitem[{{Lee} {et~al.}(2010){Lee}, {Lee}, {Park}, \& {Choi}}]{jhlee10env}
{Lee}, J.~H., {Lee}, M.~G., {Park}, C., \& {Choi}, Y.-Y. 2010, \mnras, 403,
  1930

\bibitem[{{Lewis} {et~al.}(2002){Lewis}, {Balogh}, {De Propris}, {Couch},
  {Bower}, {Offer}, {Bland-Hawthorn}, {Baldry}, {Baugh}, {Bridges}, {Cannon},
  {Cole}, {Colless}, {Collins}, {Cross}, {Dalton}, {Driver}, {Efstathiou},
  {Ellis}, {Frenk}, {Glazebrook}, {Hawkins}, {Jackson}, {Lahav}, {Lumsden},
  {Maddox}, {Madgwick}, {Norberg}, {Peacock}, {Percival}, {Peterson},
  {Sutherland}, \& {Taylor}}]{lew02}
{Lewis}, I., {Balogh}, M., {De Propris}, R., {et~al.} 2002, \mnras, 334, 673

\bibitem[{{Li} {et~al.}(2008){Li}, {Kauffmann}, {Heckman}, {Jing}, \&
  {White}}]{li08}
{Li}, C., {Kauffmann}, G., {Heckman}, T.~M., {Jing}, Y.~P., \& {White},
  S.~D.~M. 2008, \mnras, 385, 1903

\bibitem[{{Lin} {et~al.}(2007){Lin}, {Koo}, {Weiner}, {Chiueh}, {Coil}, {Lotz},
  {Conselice}, {Willner}, {Smith}, {Guhathakurta}, {Huang}, {Le Floc'h},
  {Noeske}, {Willmer}, {Cooper}, \& {Phillips}}]{lin07}
{Lin}, L., {Koo}, D.~C., {Weiner}, B.~J., {et~al.} 2007, \apjl, 660, L51

\bibitem[{{Lutz} {et~al.}(1998){Lutz}, {Spoon}, {Rigopoulou}, {Moorwood}, \&
  {Genzel}}]{lutz98}
{Lutz}, D., {Spoon}, H.~W.~W., {Rigopoulou}, D., {Moorwood}, A.~F.~M., \&
  {Genzel}, R. 1998, \apjl, 505, L103

\bibitem[{{Magdis} {et~al.}(2010{\natexlab{a}}){Magdis}, {Elbaz}, {Hwang},
  {Amblard}, {Arumugam}, {Aussel}, {Blain}, {Bock}, {Boselli}, {Buat},
  {Castro-Rodr{\'{\i}}guez}, {Cava}, {Chanial}, {Clements}, {Conley},
  {Conversi}, {Cooray}, {Dowell}, {Dwek}, {Eales}, {Farrah}, {Franceschini},
  {Glenn}, {Griffin}, {Halpern}, {Hatziminaoglou}, {Huang}, {Ibar}, {Isaak},
  {Le Floc'h}, {Lagache}, {Levenson}, {Lonsdale}, {Lu}, {Madden}, {Maffei},
  {Mainetti}, {Marchetti}, {Morrison}, {Nguyen}, {O'Halloran}, {Oliver},
  {Omont}, {Owen}, {Page}, {Pannella}, {Panuzzo}, {Papageorgiou}, {Pearson},
  {P{\'e}rez-Fournon}, {Pohlen}, {Rigopoulou}, {Rizzo}, {Roseboom},
  {Rowan-Robinson}, {Schulz}, {Scott}, {Seymour}, {Shupe}, {Smith}, {Stevens},
  {Strazzullo}, {Symeonidis}, {Trichas}, {Tugwell}, {Vaccari}, {Valtchanov},
  {Vigroux}, {Wang}, {Wright}, {Xu}, \& {Zemcov}}]{gmag10bumpy}
{Magdis}, G.~E., {Elbaz}, D., {Hwang}, H.~S., {et~al.} 2010{\natexlab{a}},
  \mnras, 409, 22

\bibitem[{{Magdis} {et~al.}(2010{\natexlab{b}}){Magdis}, {Elbaz}, {Hwang},
  {Daddi}, {Rigopoulou}, {Altieri}, {Andreani}, {Aussel}, {Berta}, {Cava},
  {Bongiovanni}, {Cepa}, {Cimatti}, {Dickinson}, {Dominguez}, {F{\"o}rster
  Schreiber}, {Genzel}, {Huang}, {Lutz}, {Maiolino}, {Magnelli}, {Morrison},
  {Nordon}, {P{\'e}rez Garc{\'{\i}}a}, {Poglitsch}, {Popesso}, {Pozzi},
  {Riguccini}, {Rodighiero}, {Saintonge}, {Santini}, {Sanchez-Portal}, {Shao},
  {Sturm}, {Tacconi}, \& {Valtchanov}}]{gmag10lbg}
{Magdis}, G.~E., {Elbaz}, D., {Hwang}, H.~S., {et~al.} 2010{\natexlab{b}},
  \apjl, 720, L185

\bibitem[{{Magdis} {et~al.}(2010{\natexlab{c}}){Magdis}, {Rigopoulou}, {Huang},
  \& {Fazio}}]{gmag10sfrm}
{Magdis}, G.~E., {Rigopoulou}, D., {Huang}, J., \& {Fazio}, G.~G.
  2010{\natexlab{c}}, \mnras, 401, 1521

\bibitem[{{Magdis} {et~al.}(2011)}]{gmag11}
{Magdis}, G.~E. {et~al.} 2011, \aap, in press (arXiv:1108.0838)

\bibitem[{{Mignoli} {et~al.}(2005){Mignoli}, {Cimatti}, {Zamorani}, {Pozzetti},
  {Daddi}, {Renzini}, {Broadhurst}, {Cristiani}, {D'Odorico}, {Fontana},
  {Giallongo}, {Gilmozzi}, {Menci}, \& {Saracco}}]{mig05}
{Mignoli}, M., {Cimatti}, A., {Zamorani}, G., {et~al.} 2005, \aap, 437, 883

\bibitem[{{Moshir} {et~al.}(1992){Moshir}, {Kopman}, \& {Conrow}}]{mos92}
{Moshir}, M., {Kopman}, G., \& {Conrow}, T.~A.~O. 1992, {IRAS Faint Source
  Survey, Explanatory supplement version 2}

\bibitem[{{Murakami} {et~al.}(2007){Murakami}, {Baba}, {Barthel}, {Clements},
  {Cohen}, {Doi}, {Enya}, {Figueredo}, {Fujishiro}, {Fujiwara}, {Fujiwara},
  {Garcia-Lario}, {Goto}, {Hasegawa}, {Hibi}, {Hirao}, {Hiromoto}, {Hong},
  {Imai}, {Ishigaki}, {Ishiguro}, {Ishihara}, {Ita}, {Jeong}, {Jeong},
  {Kaneda}, {Kataza}, {Kawada}, {Kawai}, {Kawamura}, {Kessler}, {Kester},
  {Kii}, {Kim}, {Kim}, {Kobayashi}, {Koo}, {Kwon}, {Lee}, {Lorente}, {Makiuti},
  {Matsuhara}, {Matsumoto}, {Matsuo}, {Matsuura}, {M{\"u}ller}, {Murakami},
  {Nagata}, {Nakagawa}, {Naoi}, {Narita}, {Noda}, {Oh}, {Ohnishi}, {Ohyama},
  {Okada}, {Okuda}, {Oliver}, {Onaka}, {Ootsubo}, {Oyabu}, {Pak}, {Park},
  {Pearson}, {Rowan-Robinson}, {Saito}, {Sakon}, {Salama}, {Sato}, {Savage},
  {Serjeant}, {Shibai}, {Shirahata}, {Sohn}, {Suzuki}, {Takagi}, {Takahashi},
  {Tanab{\'e}}, {Takeuchi}, {Takita}, {Thomson}, {Uemizu}, {Ueno}, {Usui},
  {Verdugo}, {Wada}, {Wang}, {Watabe}, {Watarai}, {White}, {Yamamura},
  {Yamauchi}, \& {Yasuda}}]{mur07}
{Murakami}, H., {Baba}, H., {Barthel}, P., {et~al.} 2007, \pasj, 59, 369

\bibitem[{{Nikolic} {et~al.}(2004){Nikolic}, {Cullen}, \& {Alexander}}]{nik04}
{Nikolic}, B., {Cullen}, H., \& {Alexander}, P. 2004, \mnras, 355, 874

\bibitem[{{Park} \& {Choi}(2005)}]{pc05}
{Park}, C. \& {Choi}, Y. 2005, \apjl, 635, L29

\bibitem[{{Park} \& {Choi}(2009)}]{pc09}
{Park}, C. \& {Choi}, Y. 2009, \apj, 691, 1828

\bibitem[{{Park} {et~al.}(2007){Park}, {Choi}, {Vogeley}, {Gott}, \&
  {Blanton}}]{park07}
{Park}, C., {Choi}, Y., {Vogeley}, M.~S., {Gott}, J.~R.~I., \& {Blanton}, M.~R.
  2007, \apj, 658, 898

\bibitem[{{Park} {et~al.}(2008){Park}, {Gott}, \& {Choi}}]{park08}
{Park}, C., {Gott}, J.~R.~I., \& {Choi}, Y. 2008, \apj, 674, 784

\bibitem[{{Park} \& {Hwang}(2009)}]{ph09}
{Park}, C. \& {Hwang}, H.~S. 2009, \apj, 699, 1595

\bibitem[{{Patel} {et~al.}(2009){Patel}, {Holden}, {Kelson}, {Illingworth}, \&
  {Franx}}]{pat09}
{Patel}, S.~G., {Holden}, B.~P., {Kelson}, D.~D., {Illingworth}, G.~D., \&
  {Franx}, M. 2009, \apjl, 705, L67

\bibitem[{{Patton} {et~al.}(2011){Patton}, {Ellison}, {Simard}, {McConnachie},
  \& {Mendel}}]{pat11}
{Patton}, D.~R., {Ellison}, S.~L., {Simard}, L., {McConnachie}, A.~W., \&
  {Mendel}, J.~T. 2011, \mnras, 412, 591

\bibitem[{{Patton} {et~al.}(2005){Patton}, {Grant}, {Simard}, {Pritchet},
  {Carlberg}, \& {Borne}}]{pat05}
{Patton}, D.~R., {Grant}, J.~K., {Simard}, L., {et~al.} 2005, \aj, 130, 2043

\bibitem[{{Patton} {et~al.}(1997){Patton}, {Pritchet}, {Yee}, {Ellingson}, \&
  {Carlberg}}]{pat97}
{Patton}, D.~R., {Pritchet}, C.~J., {Yee}, H.~K.~C., {Ellingson}, E., \&
  {Carlberg}, R.~G. 1997, \apj, 475, 29

\bibitem[{{Peebles}(1993)}]{pee93}
{Peebles}, P.~J.~E. 1993, {Principles of physical cosmology}, ed. P.~J.~E.
  Peebles

\bibitem[{{Perez} {et~al.}(2009){Perez}, {Tissera}, {Padilla}, {Alonso}, \&
  {Lambas}}]{perez09}
{Perez}, J., {Tissera}, P., {Padilla}, N., {Alonso}, M.~S., \& {Lambas}, D.~G.
  2009, \mnras, 399, 1157

\bibitem[{{Pilbratt} {et~al.}(2010){Pilbratt}, {Riedinger}, {Passvogel},
  {Crone}, {Doyle}, {Gageur}, {Heras}, {Jewell}, {Metcalfe}, {Ott}, \&
  {Schmidt}}]{pil10}
{Pilbratt}, G.~L., {Riedinger}, J.~R., {Passvogel}, T., {et~al.} 2010, \aap,
  518, L1

\bibitem[{{Poglitsch} {et~al.}(2010){Poglitsch}, {Waelkens}, {Geis},
  {Feuchtgruber}, {Vandenbussche}, {Rodriguez}, {Krause}, {Renotte}, {van
  Hoof}, {Saraceno}, {Cepa}, {Kerschbaum}, {Agn{\`e}se}, {Ali}, {Altieri},
  {Andreani}, {Augueres}, {Balog}, {Barl}, {Bauer}, {Belbachir}, {Benedettini},
  {Billot}, {Boulade}, {Bischof}, {Blommaert}, {Callut}, {Cara}, {Cerulli},
  {Cesarsky}, {Contursi}, {Creten}, {De Meester}, {Doublier}, {Doumayrou},
  {Duband}, {Exter}, {Genzel}, {Gillis}, {Gr{\"o}zinger}, {Henning},
  {Herreros}, {Huygen}, {Inguscio}, {Jakob}, {Jamar}, {Jean}, {de Jong},
  {Katterloher}, {Kiss}, {Klaas}, {Lemke}, {Lutz}, {Madden}, {Marquet},
  {Martignac}, {Mazy}, {Merken}, {Montfort}, {Morbidelli}, {M{\"u}ller},
  {Nielbock}, {Okumura}, {Orfei}, {Ottensamer}, {Pezzuto}, {Popesso},
  {Putzeys}, {Regibo}, {Reveret}, {Royer}, {Sauvage}, {Schreiber}, {Stegmaier},
  {Schmitt}, {Schubert}, {Sturm}, {Thiel}, {Tofani}, {Vavrek}, {Wetzstein},
  {Wieprecht}, \& {Wiezorrek}}]{pog10}
{Poglitsch}, A., {Waelkens}, C., {Geis}, N., {et~al.} 2010, \aap, 518, L2

\bibitem[{{Popesso} {et~al.}(2009){Popesso}, {Dickinson}, {Nonino}, {Vanzella},
  {Daddi}, {Fosbury}, {Kuntschner}, {Mainieri}, {Cristiani}, {Cesarsky},
  {Giavalisco}, {Renzini}, \& {GOODS Team}}]{pop09}
{Popesso}, P., {Dickinson}, M., {Nonino}, M., {et~al.} 2009, \aap, 494, 443

\bibitem[{{Popesso} {et~al.}(2011{\natexlab{a}}){Popesso}, {Rodighiero},
  {Saintonge}, {Santini}, {Grazian}, {Lutz}, {Brusa}, \& {PEP
  Consortium}}]{pop11}
{Popesso}, P., {Rodighiero}, G., {Saintonge}, A., {et~al.} 2011{\natexlab{a}},
  \aap, in press (arXiv:1104.1094)

\bibitem[{{Popesso} {et~al.}(2011{\natexlab{b}})}]{pop11mass}
{Popesso}, P. {et~al.} 2011{\natexlab{b}}, \aap, submitted

\bibitem[{{Ravikumar} {et~al.}(2007){Ravikumar}, {Puech}, {Flores}, {Proust},
  {Hammer}, {Lehnert}, {Rawat}, {Amram}, {Balkowski}, {Burgarella}, {Cassata},
  {Cesarsky}, {Cimatti}, {Combes}, {Daddi}, {Dannerbauer}, {di Serego
  Alighieri}, {Elbaz}, {Guiderdoni}, {Kembhavi}, {Liang}, {Pozzetti},
  {Vergani}, {Vernet}, {Wozniak}, \& {Zheng}}]{rav07}
{Ravikumar}, C.~D., {Puech}, M., {Flores}, H., {et~al.} 2007, \aap, 465, 1099

\bibitem[{{Reddy} {et~al.}(2006){Reddy}, {Steidel}, {Erb}, {Shapley}, \&
  {Pettini}}]{red06}
{Reddy}, N.~A., {Steidel}, C.~C., {Erb}, D.~K., {Shapley}, A.~E., \& {Pettini},
  M. 2006, \apj, 653, 1004

\bibitem[{{Retzlaff} {et~al.}(2010){Retzlaff}, {Rosati}, {Dickinson},
  {Vandame}, {Rit{\'e}}, {Nonino}, {Cesarsky}, \& {GOODS Team}}]{ret10}
{Retzlaff}, J., {Rosati}, P., {Dickinson}, M., {et~al.} 2010, \aap, 511, A50+

\bibitem[{{Rigopoulou} {et~al.}(1999){Rigopoulou}, {Spoon}, {Genzel}, {Lutz},
  {Moorwood}, \& {Tran}}]{rig99}
{Rigopoulou}, D., {Spoon}, H.~W.~W., {Genzel}, R., {et~al.} 1999, \aj, 118,
  2625

\bibitem[{{Salpeter}(1955)}]{sal55}
{Salpeter}, E.~E. 1955, \apj, 121, 161

\bibitem[{{Sanders} {et~al.}(2003){Sanders}, {Mazzarella}, {Kim}, {Surace}, \&
  {Soifer}}]{san03}
{Sanders}, D.~B., {Mazzarella}, J.~M., {Kim}, D., {Surace}, J.~A., \& {Soifer},
  B.~T. 2003, \aj, 126, 1607

\bibitem[{{Schlegel} {et~al.}(1998){Schlegel}, {Finkbeiner}, \&
  {Davis}}]{sch98}
{Schlegel}, D.~J., {Finkbeiner}, D.~P., \& {Davis}, M. 1998, \apj, 500, 525

\bibitem[{{Silverman} {et~al.}(2010){Silverman}, {Mainieri}, {Salvato},
  {Hasinger}, {Bergeron}, {Capak}, {Szokoly}, {Finoguenov}, {Gilli}, {Rosati},
  {Tozzi}, {Vignali}, {Alexander}, {Brandt}, {Lehmer}, {Luo}, {Rafferty},
  {Xue}, {Balestra}, {Bauer}, {Brusa}, {Comastri}, {Kartaltepe}, {Koekemoer},
  {Miyaji}, {Schneider}, {Treister}, {Wisotski}, \& {Schramm}}]{sil10}
{Silverman}, J.~D., {Mainieri}, V., {Salvato}, M., {et~al.} 2010, \apjs, 191,
  124

\bibitem[{{Smith} {et~al.}(2007){Smith}, {Struck}, {Hancock}, {Appleton},
  {Charmandaris}, \& {Reach}}]{smi07}
{Smith}, B.~J., {Struck}, C., {Hancock}, M., {et~al.} 2007, \aj, 133, 791

\bibitem[{{Struck}(2006)}]{str06}
{Struck}, C. 2006, {Galaxy Collisions - Dawn of a New Era}, ed. {Mason, J.~W.}
  (Springer Verlag), 115--+

\bibitem[{{Szokoly} {et~al.}(2004){Szokoly}, {Bergeron}, {Hasinger}, {Lehmann},
  {Kewley}, {Mainieri}, {Nonino}, {Rosati}, {Giacconi}, {Gilli}, {Gilmozzi},
  {Norman}, {Romaniello}, {Schreier}, {Tozzi}, {Wang}, {Zheng}, \&
  {Zirm}}]{szo04}
{Szokoly}, G.~P., {Bergeron}, J., {Hasinger}, G., {et~al.} 2004, \apjs, 155,
  271

\bibitem[{{Tegmark} {et~al.}(2004){Tegmark}, {Blanton}, {Strauss}, {Hoyle},
  {Schlegel}, {Scoccimarro}, {Vogeley}, {Weinberg}, {Zehavi}, {Berlind},
  {Budavari}, {Connolly}, {Eisenstein}, {Finkbeiner}, {Frieman}, {Gunn},
  {Hamilton}, {Hui}, {Jain}, {Johnston}, {Kent}, {Lin}, {Nakajima}, {Nichol},
  {Ostriker}, {Pope}, {Scranton}, {Seljak}, {Sheth}, {Stebbins}, {Szalay},
  {Szapudi}, {Verde}, {Xu}, {Annis}, {Bahcall}, {Brinkmann}, {Burles},
  {Castander}, {Csabai}, {Loveday}, {Doi}, {Fukugita}, {Gott}, {Hennessy},
  {Hogg}, {Ivezi{\'c}}, {Knapp}, {Lamb}, {Lee}, {Lupton}, {McKay}, {Kunszt},
  {Munn}, {O'Connell}, {Peoples}, {Pier}, {Richmond}, {Rockosi}, {Schneider},
  {Stoughton}, {Tucker}, {Vanden Berk}, {Yanny}, \& {York}}]{teg04}
{Tegmark}, M., {Blanton}, M.~R., {Strauss}, M.~A., {et~al.} 2004, \apj, 606,
  702

\bibitem[{{Telesco} {et~al.}(1988){Telesco}, {Wolstencroft}, \& {Done}}]{tel88}
{Telesco}, C.~M., {Wolstencroft}, R.~D., \& {Done}, C. 1988, \apj, 329, 174

\bibitem[{{Toomre}(1977)}]{too77}
{Toomre}, A. 1977, in Evolution of Galaxies and Stellar Populations, ed.
  {B.~M.~Tinsley \& R.~B.~Larson}, 401--+

\bibitem[{{Vanzella} {et~al.}(2008){Vanzella}, {Cristiani}, {Dickinson},
  {Giavalisco}, {Kuntschner}, {Haase}, {Nonino}, {Rosati}, {Cesarsky},
  {Ferguson}, {Fosbury}, {Grazian}, {Moustakas}, {Rettura}, {Popesso},
  {Renzini}, {Stern}, \& {GOODS Team}}]{van08}
{Vanzella}, E., {Cristiani}, S., {Dickinson}, M., {et~al.} 2008, \aap, 478, 83

\bibitem[{{Vanzella} {et~al.}(2005){Vanzella}, {Cristiani}, {Dickinson},
  {Kuntschner}, {Moustakas}, {Nonino}, {Rosati}, {Stern}, {Cesarsky}, {Ettori},
  {Ferguson}, {Fosbury}, {Giavalisco}, {Haase}, {Renzini}, {Rettura}, {Serra},
  \& {The Goods Team}}]{van05}
{Vanzella}, E., {Cristiani}, S., {Dickinson}, M., {et~al.} 2005, \aap, 434, 53

\bibitem[{{Vanzella} {et~al.}(2006){Vanzella}, {Cristiani}, {Dickinson},
  {Kuntschner}, {Nonino}, {Rettura}, {Rosati}, {Vernet}, {Cesarsky},
  {Ferguson}, {Fosbury}, {Giavalisco}, {Grazian}, {Haase}, {Moustakas},
  {Popesso}, {Renzini}, {Stern}, \& {GOODS Team}}]{van06}
{Vanzella}, E., {Cristiani}, S., {Dickinson}, M., {et~al.} 2006, \aap, 454, 423

\bibitem[{{Wang} {et~al.}(2010{\natexlab{a}}){Wang}, {Cowie}, {Barger},
  {Keenan}, \& {Ting}}]{wang10wir}
{Wang}, W.-H., {Cowie}, L.~L., {Barger}, A.~J., {Keenan}, R.~C., \& {Ting},
  H.-C. 2010{\natexlab{a}}, \apjs, 187, 251

\bibitem[{{Wang} {et~al.}(2010{\natexlab{b}}){Wang}, {Park}, {Hwang}, \&
  {Chen}}]{wang10}
{Wang}, Y., {Park}, C., {Hwang}, H.~S., \& {Chen}, X. 2010{\natexlab{b}}, \apj,
  718, 762

\bibitem[{{Wirth} {et~al.}(2004){Wirth}, {Willmer}, {Amico}, {Chaffee},
  {Goodrich}, {Kwok}, {Lyke}, {Mader}, {Tran}, {Barger}, {Cowie}, {Capak},
  {Coil}, {Cooper}, {Conrad}, {Davis}, {Faber}, {Hu}, {Koo}, {Le Mignant},
  {Newman}, \& {Songaila}}]{wir04}
{Wirth}, G.~D., {Willmer}, C.~N.~A., {Amico}, P., {et~al.} 2004, \aj, 127, 3121

\bibitem[{{Wolf} {et~al.}(2003){Wolf}, {Meisenheimer}, {Rix}, {Borch}, {Dye},
  \& {Kleinheinrich}}]{wolf03}
{Wolf}, C., {Meisenheimer}, K., {Rix}, H., {et~al.} 2003, \aap, 401, 73

\bibitem[{{Wong} {et~al.}(2011){Wong}, {Blanton}, {Burles}, {Coil}, {Cool},
  {Eisenstein}, {Moustakas}, {Zhu}, \& {Arnouts}}]{wong11}
{Wong}, K.~C., {Blanton}, M.~R., {Burles}, S.~M., {et~al.} 2011, \apj, 728, 119

\bibitem[{{Woods} \& {Geller}(2007)}]{woods07}
{Woods}, D.~F. \& {Geller}, M.~J. 2007, \aj, 134, 527

\bibitem[{{Woods} {et~al.}(2010){Woods}, {Geller}, {Kurtz}, {Westra},
  {Fabricant}, \& {Dell'Antonio}}]{woods10}
{Woods}, D.~F., {Geller}, M.~J., {Kurtz}, M.~J., {et~al.} 2010, \aj, 139, 1857

\bibitem[{{Xia} {et~al.}(2011){Xia}, {Malhotra}, {Rhoads}, {Pirzkal}, {Zheng},
  {Meurer}, {Straughn}, {Grogin}, \& {Floyd}}]{xia11}
{Xia}, L., {Malhotra}, S., {Rhoads}, J., {et~al.} 2011, \aj, 141, 64

\bibitem[{{Xilouris} {et~al.}(2004){Xilouris}, {Georgakakis}, {Misiriotis}, \&
  {Charmandaris}}]{xil04}
{Xilouris}, E.~M., {Georgakakis}, A.~E., {Misiriotis}, A., \& {Charmandaris},
  V. 2004, \mnras, 355, 57

\bibitem[{{Xu} {et~al.}(2010){Xu}, {Domingue}, {Cheng}, {Lu}, {Huang}, {Gao},
  {Mazzarella}, {Cutri}, {Sun}, \& {Surace}}]{xu10}
{Xu}, C.~K., {Domingue}, D., {Cheng}, Y., {et~al.} 2010, \apj, 713, 330

\bibitem[{{Yee} \& {Ellingson}(1995)}]{ye95}
{Yee}, H.~K.~C. \& {Ellingson}, E. 1995, \apj, 445, 37

\bibitem[{{York} {et~al.}(2000){York}, {Adelman}, {Anderson}, {Anderson},
  {Annis}, {Bahcall}, {Bakken}, {Barkhouser}, {Bastian}, {Berman}, {Boroski},
  {Bracker}, {Briegel}, {Briggs}, {Brinkmann}, {Brunner}, {Burles}, {Carey},
  {Carr}, {Castander}, {Chen}, {Colestock}, {Connolly}, {Crocker}, {Csabai},
  {Czarapata}, {Davis}, {Doi}, {Dombeck}, {Eisenstein}, {Ellman}, {Elms},
  {Evans}, {Fan}, {Federwitz}, {Fiscelli}, {Friedman}, {Frieman}, {Fukugita},
  {Gillespie}, {Gunn}, {Gurbani}, {de Haas}, {Haldeman}, {Harris}, {Hayes},
  {Heckman}, {Hennessy}, {Hindsley}, {Holm}, {Holmgren}, {Huang}, {Hull},
  {Husby}, {Ichikawa}, {Ichikawa}, {Ivezi{\'c}}, {Kent}, {Kim}, {Kinney},
  {Klaene}, {Kleinman}, {Kleinman}, {Knapp}, {Korienek}, {Kron}, {Kunszt},
  {Lamb}, {Lee}, {Leger}, {Limmongkol}, {Lindenmeyer}, {Long}, {Loomis},
  {Loveday}, {Lucinio}, {Lupton}, {MacKinnon}, {Mannery}, {Mantsch}, {Margon},
  {McGehee}, {McKay}, {Meiksin}, {Merelli}, {Monet}, {Munn}, {Narayanan},
  {Nash}, {Neilsen}, {Neswold}, {Newberg}, {Nichol}, {Nicinski}, {Nonino},
  {Okada}, {Okamura}, {Ostriker}, {Owen}, {Pauls}, {Peoples}, {Peterson},
  {Petravick}, {Pier}, {Pope}, {Pordes}, {Prosapio}, {Rechenmacher}, {Quinn},
  {Richards}, {Richmond}, {Rivetta}, {Rockosi}, {Ruthmansdorfer}, {Sandford},
  {Schlegel}, {Schneider}, {Sekiguchi}, {Sergey}, {Shimasaku}, {Siegmund},
  {Smee}, {Smith}, {Snedden}, {Stone}, {Stoughton}, {Strauss}, {Stubbs},
  {SubbaRao}, {Szalay}, {Szapudi}, {Szokoly}, {Thakar}, {Tremonti}, {Tucker},
  {Uomoto}, {Vanden Berk}, {Vogeley}, {Waddell}, {Wang}, {Watanabe},
  {Weinberg}, {Yanny}, \& {Yasuda}}]{york00}
{York}, D.~G., {Adelman}, J., {Anderson}, Jr., J.~E., {et~al.} 2000, \aj, 120,
  1579

\bibitem[{{Zepf} \& {Koo}(1989)}]{zk89}
{Zepf}, S.~E. \& {Koo}, D.~C. 1989, \apj, 337, 34

\end{thebibliography}
\end{document}